\documentclass[aps,prd,11pt,superscriptaddress,notitlepage,longbibliography,nofootinbib,tightenlines,preprintnumbers]{revtex4-1}

\usepackage{amsmath,amssymb,amsfonts}
\usepackage{graphicx}
\usepackage{color}
\usepackage{tikz}
\usetikzlibrary{calc}
\usepackage{dsfont}
\usepackage[pdftex]{hyperref}
\hypersetup{colorlinks=true, linkcolor=darkred, citecolor=blue, linktoc=page}
\definecolor{darkred}{rgb}{0.8,0.1,0.1}

\numberwithin{equation}{section}
\renewcommand\theequation{\arabic{section}.\arabic{equation}} 

\usepackage{subfigure}

\def\cA{{\cal A}}
\def\cB{{\cal B}}
\def\cC{{\cal C}}
\def\cF{{\cal F}}
\def\cI{{\cal I}}

\def\cG{{\cal G}}
\def\cL{{\cal L}}

\def\cW{{\cal W}}
\def\cY{{\cal Y}}
\def\cZ{{\cal Z}}

\def\CC{\ensuremath{\mathbb C}}
\def\RR{\ensuremath{\mathbb R}}
\def\ZZ{\ensuremath{\mathbb Z}}

\def\fC{\ensuremath{\mathfrak{C}}}

\DeclareMathOperator{\vol}{vol}

\DeclareMathOperator{\Li}{Li}

\def\Im{\mathop{\rm Im}}
\def\Re{\mathop{\rm Re}}

\newcommand{\vertx}{\ensuremath {{\sf I}}}
\newcommand{\perpx}{{\boldmath $\perp$}}

\newcommand{\nocontentsline}[3]{}
\newcommand{\tocless}[2]{\bgroup\let\addcontentsline=\nocontentsline#1{#2}\egroup}

\begin{document}

\title{\texorpdfstring{AdS$_6$/CFT$_5$}{AdS6/CFT5} with O7-planes}

\author{Christoph F.~Uhlemann} 
\email{uhlemann@umich.edu}
  
\affiliation{Leinweber Center for Theoretical Physics, Department of Physics
	\\
	University of Michigan, Ann Arbor, MI 48109-1040, USA\\[2mm]}

\affiliation{Mani L.\ Bhaumik Institute for Theoretical Physics,
 Department of Physics and Astronomy\\
 University of California, Los Angeles, CA 90095, USA}
 
\preprint{LCTP-19-34}

\begin{abstract}
Type IIB AdS$_6$ solutions with orientifold 7-planes are constructed.
The geometry is a warped product of AdS$_6$ and S$^2$ over a Riemann surface $\Sigma$ and the O7-planes correspond to a particular type of puncture on $\Sigma$.
The solutions are identified as near-horizon limits of $(p,q)$ 5-brane webs with O7-planes.
The dual 5d SCFTs have relevant deformations to linear quiver gauge theories which have $SO(\cdot)$ or $USp(\cdot)$ nodes or $SU(\cdot)$ nodes with hypermultiplets in symmetric or antisymmetric representations, in addition to $SU(\cdot)$ nodes with fundamental hypermultiplets.
The S$^5$ free energies are obtained holographically and matched to field theory computations using supersymmetric localization to support the proposed dualities.
\end{abstract}

\maketitle

\tableofcontents

\section{Introduction}

One of the striking lessons from string theory is the existence of UV-complete interacting quantum field theories in 5 and 6 dimensions. 
These theories are interesting in their own right and provide intriguing new perspectives on lower-dimensional theories.
The focus of this work are 5-dimensional superconformal field theories (SCFTs). 
These theories are intrinsically strongly coupled and have no conventional Lagrangian description,
but they often have relevant deformations that are described by perturbatively non-renormalizable gauge theories.
The study of 5d SCFTs is facilitated by a combination of various different methods, including brane constructions and decoupling limits in string theory and M-theory, AdS/CFT dualities, and field theory studies using supersymmetric localization in gauge theory deformations. Recent studies include \cite{Jefferson:2017ahm,Jefferson:2018irk,Bhardwaj:2018yhy,Bhardwaj:2018vuu,Apruzzi:2019opn,Bhardwaj:2019xeg,Saxena:2019wuy,Penin:2019jlf,Hosseini:2019and,Apruzzi:2019kgb,Kim:2019fsg,Kim:2019uqw,Choi:2019miv,Closset:2018bjz,Cabrera:2018jxt,Crichigno:2018adf,Hosseini:2018uzp}.

Large classes of 5d SCFTs and their relevant deformations can be engineered by $(p,q)$ 5-brane webs in Type IIB string theory \cite{Aharony:1997ju,Kol:1997fv,Aharony:1997bh}, and by 5-brane webs with $[p,q]$ 7-branes and further extensions \cite{DeWolfe:1999hj,Bergman:2015dpa,Zafrir:2015ftn,Hayashi:2018bkd,Hayashi:2018lyv,Hayashi:2019yxj}.
Type IIB Supergravity solutions describing the near-horizon limit of 5-brane webs were constructed in \cite{DHoker:2016ujz,DHoker:2016ysh,DHoker:2017mds}, and solutions describing 5-brane webs with mutually local $[p,q]$ 7-branes inside the web were constructed in \cite{DHoker:2017zwj}.\footnote{Earlier studies of the BPS equations can be found in \cite{Apruzzi:2014qva,Kim:2015hya,Kim:2016rhs}. T-duals of the Brandhuber/Oz solution to massive Type IIA \cite{Brandhuber:1999np} were discussed in \cite{Lozano:2012au,Lozano:2013oma} and interpreted as solutions with smeared branes in \cite{Lozano:2018pcp,Lozano:2018fvt}.}
These supergravity solutions provide holographic duals for the 5d SCFTs engineered by the corresponding 5-brane webs in the conformal limit and offer new ways to study them.
Consistent truncations based on these Type IIB solutions to 6d gauged supergravity were constructed in \cite{Hong:2018amk,Malek:2019ucd,Malek:2018zcz}.

The aim of this work is to extend the class of dualities between 5d SCFTs engineered by 5-brane webs and $AdS_6$ solutions in Type IIB supergravity.
The 5-brane web constructions can be extended in interesting ways by incorporating O7 planes \cite{Bergman:2015dpa}. This allows to realize 5d SCFTs with gauge theory deformations that involve  $USp(\cdot)$ and $SO(\cdot)$ gauge groups, as well as $SU(\cdot)$ gauge groups with matter hypermultiplets in symmetric and antisymmetric representations, including theories that were excluded in the initial classification \cite{Intriligator:1997pq}. 
In this paper the construction of $AdS_6$ supergravity solutions will be extended to incorporate O7-planes.
The solutions are constructed as quotients of certain limiting cases of the solutions with $[p,q]$ 7-branes constructed in \cite{DHoker:2017zwj}, for which new explicit expressions are derived.
The resulting solutions with O7-planes  will be identified as near-horizon limits of 5-brane webs that involve O7-planes localized inside the web, and as holographic duals for the 5d SCFTs engineered by such 5-brane webs.

5d SCFTs with holographic duals in Type IIB often have relevant deformations that are described by gauge theories in the IR.
For the theories corresponding to 5-brane webs with D7-branes these are linear quiver theories with $SU(\cdot)$  nodes connected by bifundamental hypermultiplets, possibly involving Chern-Simons terms and additional fundamental hypermultiplets.
The $S^5$ free energies for such theories were matched between supergravity and field theory computations in \cite{Fluder:2018chf,Fluder:2019szh} and in \cite{Uhlemann:2019ypp}, where the general form of the quivers is discussed in detail.
The 5d SCFTs dual to the supergravity solutions with localized O7-planes have deformations to linear quiver gauge theories of the aforementioned type, but, depending on the type of O7 and whether or not fractional NS5-branes are involved,
isolated nodes in addition have hypermultiplets in symmetric or antisymmetric representations or are replaced by $SO(\cdot)$ or $USp(\cdot)$ nodes.
The $S^5$ free energies will again be matched between supergravity and field theory computations.

The remaining parts are organized as follows. In sec.~\ref{sec:AdS6-O7-sol} the construction of AdS$_6$ solutions with O7-planes is detailed. 
In sec.~\ref{sec:5dSCFTs} the identification as near-horizon limits of 5-brane webs with O7-planes is discussed.
Examples of brane junctions, 5d SCFTs, and associated supergravity solutions are discussed explicitly, 
and the free energies are computed holographically.
In sec.~\ref{sec:F-loc} the $S^5$ free energies are analyzed in field theory.
The results are discussed in sec.~\ref{sec:discussion}.

\section{\texorpdfstring{$AdS_6$}{AdS6} solutions with O7 planes}\label{sec:AdS6-O7-sol}

In this section $AdS_6$ solutions with O7 planes are constructed.
The construction is based on the general local form of supersymmetric AdS$_6$ solutions in Type IIB supergravity derived in \cite{DHoker:2016ujz}. The geometry is a warped product of $AdS_6$ and $S^2$ over a Riemann surface $\Sigma$.
The solutions are defined by a pair of locally holomorphic functions $\cA_\pm$ on $\Sigma$, from which composite quantities $\kappa^2$, $\cG$, $T$ and $R$ are defined as follows,
\begin{align}\label{eq:kappa2-G}
 \kappa^2&=-|\partial_w \cA_+|^2+|\partial_w \cA_-|^2~,
 &
 \partial_w\cB&=\cA_+\partial_w \cA_- - \cA_-\partial_w\cA_+~,
 \nonumber\\
 \cG&=|\cA_+|^2-|\cA_-|^2+\cB+\bar{\cB}~,
 &
  T^2&=\left(\frac{1+R}{1-R}\right)^2=1+\frac{2|\partial_w\cG|^2}{3\kappa^2 \, \cG }~,
\end{align}
where $w$ is a complex coordinate on $\Sigma$. The Einstein-frame metric, complex two-form $C_{(2)}$ and axion-dilaton scalar $B=(1+i\tau)/(1-i\tau)$ are then given by
\begin{align}\label{eqn:ansatz}
 ds^2 &= \sqrt{6\cG T} \, ds^2 _{\mathrm{AdS}_6} + \frac{1}{9}\sqrt{6\cG}\,T ^{-\tfrac{3}{2}} \, ds^2 _{\mathrm{S}^2} 
 + \frac{4\kappa^2}{\sqrt{6\cG}} T^{\tfrac{1}{2}}\, |dw|^2~,
\nonumber\\
 C_{(2)}&=\frac{2i}{3}\left(
 \frac{\partial_{\bar w}\cG\partial_w\cA_++\partial_w \cG \partial_{\bar w}\bar\cA_-}{3\kappa^{2}T^2} - \bar{\mathcal{A}}_{-} - \mathcal{A}_{+}  \right) \vol_{S^2}~,
 \nonumber\\
 B &=\frac{\partial_w \cA_+ \,  \partial_{\bar w} \cG - R \, \partial_{\bar w} \bar \cA_-   \partial_w \cG}{
R \, \partial_{\bar w}  \bar \cA_+ \partial_w \cG - \partial_w \cA_- \partial_{\bar w}  \cG}~,
\end{align}
where  $\vol_{S^2}$ and $ds^2_{S^2}$ are, respectively, the volume form and line element for a unit-radius $S^2$, and $ds^2_{AdS_6}$ is the line element of unit-radius $AdS_6$.
The general form of $\cA_\pm$ for physically regular solutions that involve $(p,q)$ 5-branes and mutually local 7-branes was constructed in \cite{DHoker:2016ysh,DHoker:2017mds,DHoker:2017zwj}, and the residual regularity conditions were derived.
For these solutions $\Sigma$ is a disc. The differentials $\partial_w\cA_\pm$ have poles on the boundary of the disc which correspond to 5-branes with charges given by the residues, and in the interior of the disc there can be punctures with $SL(2,\RR)$ monodromy, which correspond to 7-branes with the given monodromy. The $SL(2,\RR)$ transformations of the Type IIB supergravity fields are induced by $SU(1,1)\otimes\CC$ transformations of $\cA_\pm$. Namely,
\begin{align}\label{eq:cApm-SU11}
 \begin{pmatrix}
  \cA_+\\ \cA_-
 \end{pmatrix}
&\rightarrow
\begin{pmatrix}
  \cA_+^\prime\\ \cA_-^\prime
 \end{pmatrix}
 =
\begin{pmatrix}
  u & -v \\ 
  -\bar v & \bar u
 \end{pmatrix}
\begin{pmatrix}
  \cA_+\\ \cA_-
 \end{pmatrix}
 +\begin{pmatrix} a_0 \\ \bar a_0\end{pmatrix}
 ~.
\end{align}
The $SL(2,\RR)$ paramters with $ad-bc=1$ are related to the $SU(1,1)$ parameters by $u=\frac{1}{2}(a+d+ib-ic)$ and $v=\frac{1}{2}(-a+d+ib+ic)$.
The constant shifts affect a gauge transformation of the complex two-form.
More details can be found in sec.~2.3 of \cite{DHoker:2017zwj}.

Localized O7-planes correspond to a particular type of puncture on the Riemann surface.
The general strategy for the construction of solutions with O7-planes from solutions with D7-branes is discussed in sec.~\ref{sec:strat}.
In sec.~\ref{sec:sol-D7} explicit expressions for $\cA_\pm$ for solutions with 5-brane poles and D7-brane punctures are presented.
In sec.~\ref{sec:O7-sol} solutions with a $\ZZ_2$ symmetry are discussed, 
and in sec.~\ref{sec:sol} solutions with an O7-plane puncture are realized as quotients of solutions with D7-branes.
The characteristics of the O7 punctures are summarized in sec.~\ref{sec:near-puncture}.

\begin{figure}
\centering
\subfigure[][]{\label{fig:strategy-a}
\begin{tikzpicture}
\draw[fill=lightgray!20] (0,0) circle (1.5);
\draw[thick] (0,0) circle (1.5);
\node at (0.55,0.8) {$\mathbf \Sigma, z$};

\foreach \j in {-70,-40,10,60}{
 \draw[very thick] ({sin(\j)*1.4},{cos(\j)*1.4}) -- ({sin(\j)*1.6},{cos(\j)*1.6});
 \draw[very thick] ({sin(180+\j)*1.4},{cos(180+\j)*1.4}) -- ({sin(180+\j)*1.6},{cos(180+\j)*1.6}); 
}
\node at ({sin(60)*1.85},{cos(60)*1.85}) {$r_1$};
\node at ({sin(10)*1.8},{cos(10)*1.8}) {$r_2$}; 
\node[rotate=35] at ({sin(-40)*1.8},{cos(-40)*1.8}) {$\dots$}; 
\node at ({sin(-70)*1.9},{cos(-70)*1.9}) {$r^{}_{L_0}$}; 

\node at ({sin(180+60)*1.9},{cos(180+60)*1.9}) {$r_{1+L_0}$};
\node at ({sin(180+10)*1.8},{cos(180+10)*1.8}) {$r_{2+L_0}$}; 
\node[rotate=35] at ({sin(180-40)*1.8},{cos(180-40)*1.8}) {$\dots$}; 
\node at ({sin(180-70)*2},{cos(180-70)*2}) {$r^{}_{2L_0}$}; 

\draw[fill=black] (0.5,0) circle (0.06);
\draw[fill=black] (-0.5,0) circle (0.06);
\draw[thick,dashed, black] (0.5,0) -- (1.5,0);
\draw[thick,dashed, black] (-0.5,0) -- (-1.5,0);
\node at (0.7,-0.3) {\footnotesize $T^k$};
\node at (-0.7,-0.3) {\footnotesize $T^k$};
\end{tikzpicture}
}\hskip 10mm
\subfigure[][]{\label{fig:strategy-b}
\begin{tikzpicture}
\draw[fill=lightgray!20] (0,0) circle (1.5);
\draw[thick] (0,0) circle (1.5);
\node at (0.55,0.8) {$\mathbf \Sigma,z$};

\foreach \j in {-70,-40,10,60}{
 \draw[very thick] ({sin(\j)*1.4},{cos(\j)*1.4}) -- ({sin(\j)*1.6},{cos(\j)*1.6});
 \draw[very thick] ({sin(180+\j)*1.4},{cos(180+\j)*1.4}) -- ({sin(180+\j)*1.6},{cos(180+\j)*1.6}); 
}
\node at ({sin(60)*1.85},{cos(60)*1.85}) {$r_1$};
\node at ({sin(10)*1.8},{cos(10)*1.8}) {$r_2$}; 
\node[rotate=35] at ({sin(-40)*1.8},{cos(-40)*1.8}) {$\dots$}; 
\node at ({sin(-70)*1.9},{cos(-70)*1.9}) {$r^{}_{L_0}$}; 

\node at ({sin(180+60)*1.9},{cos(180+60)*1.9}) {$r_{1+L_0}$};
\node at ({sin(180+10)*1.8},{cos(180+10)*1.8}) {$r_{2+L_0}$}; 
\node[rotate=35] at ({sin(180-40)*1.8},{cos(180-40)*1.8}) {$\dots$}; 
\node at ({sin(180-70)*2},{cos(180-70)*2}) {$r^{}_{2L_0}$}; 

\draw[fill=black] (0,0) circle (0.08);
\draw[thick,dashed, black] (0,0) -- (1.5,0);
\draw[thick,dashed, black] (0,0) -- (-1.5,0);
\node at (0.7,-0.3) {\footnotesize $T^k$};
\node at (-0.7,-0.3) {\footnotesize $T^k$};
\end{tikzpicture}
}\hskip 10mm
\subfigure[][]{\label{fig:strategy-c}
\begin{tikzpicture}
\draw[fill=lightgray!20] (0,0) circle (1.5);
\draw[thick] (0,0) circle (1.5);
\node at (0.55,0.8) {$\mathbf \Sigma,u$};

\foreach \j in {-140,-80,20,120}{
 \draw[very thick] ({sin(\j)*1.4},{cos(\j)*1.4}) -- ({sin(\j)*1.6},{cos(\j)*1.6});
}
\node at ({sin(120)*1.95},{cos(120)*1.95}) {$r^{}_{L_0}$};
\node at ({sin(20)*1.8},{cos(20)*1.8}) {$r_1$}; 
\node[rotate=-40] at ({sin(-140)*1.8},{cos(-140)*1.8}) {$\dots$}; 
\node at ({sin(-80)*1.85},{cos(-80)*1.85}) {$r_2$}; 

\draw[very thick] (-0.1,0.1) -- (0.1,-0.1);
\draw[very thick] (-0.1,-0.1) -- (0.1,0.1);

\draw[thick,dashed, black] (0,0) -- (1.5,0);
\node at (0.7,-0.3) {\footnotesize $-T^k$};

\node at ({sin(180+10)*1.8},{cos(180+10)*1.8}) {\hphantom{$r_{2+L_0}$}}; 
\end{tikzpicture}
} 

 \caption{On the left a supergravity solution with two punctures and an even number of poles such that a rotation by $\pi$ combined with an $SL(2,\ZZ)$ transformation by $-\mathds{1}_2$ maps the solution to itself.
 In the center the limit where the punctures become coincident. 
 Taking a quotient by the combined rotation and $SL(2,\ZZ)$ transformation that is a symmetry of the solution leads to the solution illustrated on the right. The two branch cuts are identified and the combined monodromy becomes $-T^k$.
 \label{fig:strategy}}
\end{figure}
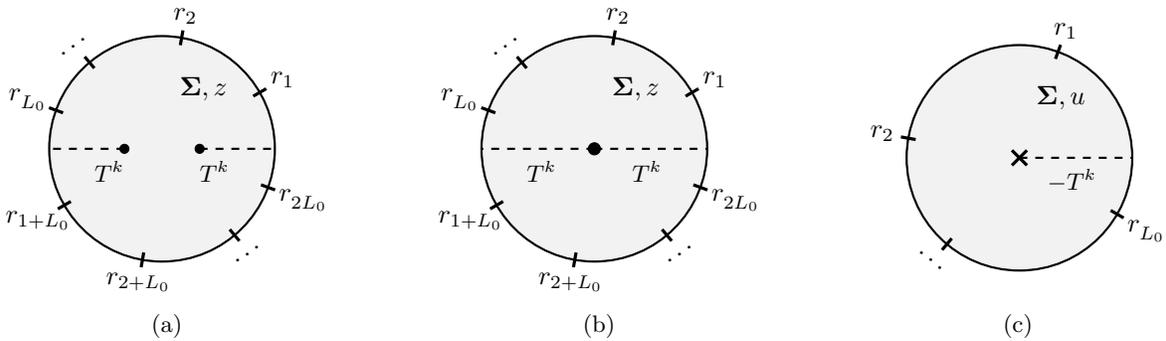

\subsection{Strategy}\label{sec:strat}

In the $AdS_6$ supergravity solutions 7-branes are naturally realized as punctures on the Riemann surface $\Sigma$ around which the supergravity fields undergo the $SL(2,\RR)$ monodromy characterizing the 7-brane. 
O7$^\pm$ planes are characterized by $SL(2,\RR)$ monodromies
\begin{align}\label{eq:MO7}
 M_{\rm O7^\pm}&=-T^{\pm 4}~, & T&=\begin{pmatrix} 1 & 1 \\ 0 & 1\end{pmatrix}~,
\end{align}
where $T$ is the monodromy of a D7-brane.
The monodromy for an O7$^-$ with $8$ D7 branes is identical to that of an O7$^+$.
More generally, a combination of an O7$^+$ with $n$ D7-branes is characterized by the same monodromy as an O7$^-$ with $n+8$ D7-branes.
The two configurations have different deformations\footnote{%
In F-theory the two types of orientifolds correspond to the frozen and unfrozen phases \cite{Tachikawa:2015wka,Bhardwaj:2018jgp}.}
and the action of the orientifolds on open strings is different,
but the supergravity solutions are expected to be identical.

The starting point for realizing solutions with punctures of monodromy $-T^k$ are solutions with 5-brane poles and D7-brane punctures which have a $\ZZ_2$ symmetry.
The solutions are represented with $\Sigma$ realized as a disc, as in fig.~\ref{fig:strategy-a}.
They have an even number $L=2L_0$ of 5-brane poles in the differentials $\partial_w\cA_\pm$,
which are arranged in pairs symmetrically under reflection across the origin, at $(r_\ell,r_{\ell+L_0})$.
The residues within each pair are opposite-equal.
In addition there may be pairs of D7-brane punctures with associated branch cuts, also arranged symmetrically under reflection across the origin.
One pair of punctures, each with monodromy $T^k$, is on a horizontal diameter of the disc with their branch cuts pointing outwards. 
The case $k=0$, corresponding to vanishing D7-brane charge and no punctures, is allowed.
A solution of this type is mapped into itself by a rotation of the disc by $\pi$ combined with a sign flip for all residues.
With $z$ a complex coordinate on the disc the combined $\ZZ_2$ action is
\begin{align}\label{eq:fC-disc}
 \fC:&& z&\rightarrow -z & \partial_z\cA_\pm & \rightarrow -\partial_z\cA_\pm~.
\end{align}
The action on $\partial\cA_\pm$ corresponds to the element $-\mathds{1}_2\in SL(2,\RR)$, as seen from eq.~(\ref{eq:cApm-SU11}).

Consider now the limit where the D7-brane punctures on the horizontal diameter are moved inwards towards the center of the disc. The limit where they have become coincident is shown in fig.~\ref{fig:strategy-b}. The D7-branes are at the fixed point of the $\fC$ transformation, and their branch cuts are mapped into each other.
If the limit is taken within a family of regular solutions with separated punctures, the solution with coincident punctures will satisfy the regularity conditions as well.
The orientifold projection can then be realized as a quotient by $\fC$, which corresponds to considering one half of the disc with the two ``half branch cuts'' along the horizontal diameter identified.
A convenient coordinate on the fundamental domain of the quotient solution is
\begin{align}\label{eq:z-u}
 z&=i\sqrt{-u}~,
\end{align}
such that $u$ runs through a unit disc. The solution in the $u$ coordinate is shown in fig.~\ref{fig:strategy-c}, with the branch cut along the positive real axis.
Crossing the branch cut in the quotient solution corresponds to crossing a half branch cut in the parent solution followed by a $\fC$ transformation, such that the monodromy in the quotient solution is given by
\begin{align}\label{eq:monodromy}
 T^k\cdot (-\mathds{1}_2)&=-T^k=\begin{pmatrix} -1 & -k \\ 0 & -1\end{pmatrix}~.
\end{align}
The parent solution requires $k \geq 0$ to be regular, as discussed in \cite{DHoker:2017zwj}.
For generic $k\geq 0$, (\ref{eq:monodromy}) is the appropriate monodromy for an O7$^-$ with $4+k$ D7-branes. 
For $k=0$ the parent solution has no puncture at the center, and the puncture in the quotient solution corresponds to an O7$^0$.
For $k\geq 4$ the monodromy can also be associated with an O7$^+$ with $k-4$ D7-branes.
In particular, for $k=4$ it can be associated with an isolated O7$^+$.

\subsection{Solutions with D7-branes}\label{sec:sol-D7}

The starting point for the strategy outlined in sec.~\ref{sec:strat} are $AdS_6$ solutions with D7-brane punctures.
Such solutions were constructed in \cite{DHoker:2017zwj}. Explicit expressions were derived for the differentials $\partial\cA_\pm$; the functions $\cA_\pm$ and the regularity conditions were given in integral representations.

Explicit expressions for the general form of $\cA_\pm$ and the residual regularity conditions are derived in appendix \ref{app:Apm-expl}. 
They are expressed in terms of the function
\begin{align}\label{eq:G2-def}
 L(z,t)&=\Li_2\left(\frac{z}{t}\right)+\ln(z)\ln\left(1-\frac{z}{t}\right)~.
\end{align}
It has two branch cuts, one starting at $z=0$ and extending along the negative real axis, and one starting at $z=t$ and extending along $z=\alpha t$ with $\alpha\geq 1$.
For $\Sigma$ the upper half plane with complex coordinate $w$, the functions $\cA_\pm$ for solutions with D7-brane punctures at $w_i$ and 5-brane poles at $r_\ell$ can then be written as
\begin{align}
\cA_\pm(w)&=\cA_\pm^0 +  \sum_{\ell=1}^L  \left[Z_\pm^\ell\ln(w-r_\ell)
+Y^\ell\sum _{i=1}^I \frac{n_i^2}{4\pi}L\left(\gamma_i\frac{w-w_i}{w-\bar w_i},\gamma_i\frac{r_\ell-w_i}{r_\ell-\bar w_i}\right)\right]~,
\label{eq:cApm-D7}
\end{align}
where $Y^\ell\equiv Z_+^\ell-Z_-^\ell$. The parameters are constrained by $\overline{Z_\pm^\ell}=-Z_\mp^\ell$ and $\sum_{\ell=1}^L Z_\pm^\ell=0$.
The orientation of the branch cuts associated with the punctures is parametrized by phases $\gamma_i$, and the monodromies are $T^{n_i^2}$. 
The residues of $\partial_w\cA_\pm$ at the poles $r_\ell$ are 
\begin{align}\label{eq:cY}
\cY_\pm^\ell&\equiv Z_\pm^\ell+ Y^\ell\sum_{i=1}^I\frac{n_i^2}{4\pi}\ln\left(\gamma_i\frac{r_\ell-w_i}{r_\ell-\bar w_i}\right)~.
\end{align}
The residues encode the 5-brane charges $(p_\ell,q_\ell)$ at the pole, with $(1,0)$ corresponding to a D5-brane and $(0,1)$ corresponding to an NS5 brane, via \cite{Bergman:2018hin}
\begin{align}\label{eq:cY-pq}
 \cY_+^\ell&=\frac{3}{4}\alpha^\prime \left(q_\ell+i p_\ell\right)~.
\end{align}
The residual regularity conditions are given by
\begin{subequations}\label{eq:reg-D7}
\begin{align}
  0&=2\cA_+^0-2\cA_-^0+\sum_{\ell=1}^LY^\ell \ln|w_i-r_\ell|^2~, & i&=1,\ldots,I~,
  \label{eq:reg-D7-1}\\
  0&=2\cA_+^0Z_-^k-2\cA_-^0Z_+^k+\sum_{\ell\neq k}\left[
  Z^{[\ell,k]}\ln|r_\ell-r_k|^2
  -Y^\ell Y^k (\cI_{k\ell}-\cI_{\ell k})
  \right],
  & k&=1,\ldots L~,
  \label{eq:reg-D7-2}
\end{align}
\end{subequations}
with
\begin{align}\label{eq:cI-def}
 \cI_{k\ell}&=\sum_{i=1}^I\frac{n_i^2}{4\pi}
  \Li_2\left(\frac{r_k-w_i}{r_k-\bar w_i}\frac{r_\ell-\bar w_i}{r_\ell-w_i}\right)~.
\end{align}
Note that $\cI_{k\ell}-\cI_{\ell k}$ is imaginary and single valued, and the sum of the conditions in (\ref{eq:reg-D7-2}) vanishes.
For a non-degenerate solution with  D7-brane punctures $L\geq 2$ \cite{Chaney:2018gjc}, otherwise $L\geq 3$ \cite{DHoker:2017mds}.

\subsection{Solutions with \texorpdfstring{$\ZZ_2$}{Z2} symmetry}\label{sec:O7-sol}

The next step is to implement the $\ZZ_2$ symmetry (\ref{eq:fC-disc}) on the upper half plane.
Two of the degrees of freedom in the $SL(2,\RR)$ automorphisms of the upper half plane can be used to realize the fixed point at $w=i$.
The transformation (\ref{eq:fC-disc}) is then implemented by
\begin{align}\label{eq:fC}
 \fC: &&
 w&\rightarrow -\frac{1}{w}~, & 
 \partial\cA_\pm&\rightarrow -\partial\cA_\pm~.
\end{align}
To realize an even number of poles, with residues related in pairs by charge conjugation, start with
\begin{align}\label{eq:sol-O7-1}
 L&=2L_0~, &  Z_\pm^\ell=-Z_\pm ^{\ell+L_0}~, \qquad \ell=1,..,L_0~.
\end{align}
The positions of the poles are chosen symmetrically with respect to $w\rightarrow -1/w$,
without loss of generality such that $r_1,\ldots,r_{L_0}$ are non-negative,
\begin{align}\label{eq:sol-O7-1a}
 r_{\ell+L_0}&=-\frac{1}{r_\ell}~, \qquad \ell=1,\ldots,L_0~,
 &
 r_1,\ldots,r_{L_0}\geq 0~.
\end{align}
The $\ZZ_2$ symmetry allows for an even number of punctures, with positions and branch cuts symmetric under reflection across the imaginary axis,\footnote{It is also assumed that the $\gamma_i$ are chosen such that the branch cuts associated with punctures in the right/left quadrant intersect the positive/negative real line.}
\begin{align}\label{eq:sol-O7-2}
 I&=2I_0~, & w_{i+I_0}&=-\frac{1}{w_i}~, & n_{i+I_0}&=n_i~, & \gamma_{i+I_0}&=\frac{w_i}{\bar w_i}\gamma_i~, & i&=1,\ldots I_0~.
\end{align}
Without loss of generality the punctures are ordered such that $w_1,\ldots w_{I_0}$ are in the right quadrant of the upper half plane or on the imaginary axis.
With the $Z_\pm^\ell$ related as in (\ref{eq:sol-O7-1}), (\ref{eq:cY}) yields
\begin{align}
 \cY_\pm^\ell=-\cY_\pm ^{\ell+L_0}~, \qquad \ell=1,\ldots,L_0~.
\end{align}
Since the $\cY_\pm^\ell$ encode the 5-brane charges, this shows that the parameter choice realizes the symmetry discussed in sec.~\ref{sec:strat}.

The explicit expressions for $\cA_\pm$ are obtained straightforwardly from (\ref{eq:cApm-D7}).
As shown in app.~\ref{eq:app-O7}, the regularity conditions fix $\cA_\pm^0$ such that
\begin{align}
 \cA_\pm(w)=\,& \sum_{\ell=1}^{L_0}\left[Z_\pm^\ell \ln\left(-i\frac{w-r_\ell}{wr_\ell+1}\right)+Y^\ell \sum _{i=1}^{I_0} \frac{n_i^2}{4\pi} \left(\tilde L_{i\ell}(w)-\tilde L_{i\ell}\left(-w^{-1}\right)\right)\right]~,
 \nonumber\\
 \tilde L_{i\ell}(w)=\,&
L\left(\gamma_i\frac{w-w_i}{w-\bar w_i},\gamma_i\frac{r_\ell-w_i}{r_\ell-\bar w_i}\right)
+L\left(\gamma_i\frac{w w_i+1}{w \bar w_i+1},\gamma_i\frac{r_\ell w_i+1}{r_\ell \bar w_i+1}\right)
~.
\label{eq:Apm-Z2}
\end{align}
The $2L_0+2I_0$ regularity conditions in (\ref{eq:reg-D7}) reduce to half, as expected based on the symmetry of the setup and shown in detail in app.~\ref{eq:app-O7}. 
The remaining regularity conditions are
\begin{subequations}\label{eq:O7-reg}
\begin{align}
 0&=\sum_{\ell=1}^{L_0} Y^\ell \ln\left|\frac{w_i-r_\ell}{w_i r_\ell+1}\right|^2~, & i=1,\ldots,I_0~,
 \label{eq:O7-reg-1}
 \\
 0&=\sum_{\stackrel{\ell=1}{\ell\neq k}}^{L_0}\left[ Z^{[\ell,k]}\ln\left|\frac{r_\ell-r_k}{r_kr_\ell+1}\right|^2
 +Y^k Y^\ell\left(\tilde\cI_{k\ell}-\tilde\cI_{\ell k}\right)\right]~, & k=1,\ldots,L_0~,
 \label{eq:O7-reg-2}
\end{align}
\end{subequations}
where
\begin{align}
 \tilde\cI_{k\ell}=
 \sum_{i=1}^{I_0}\frac{n_i^2}{4\pi}\Bigg[&
 \Li_2\left(\frac{r_k-w_i}{r_k-\bar w_i}\frac{r_\ell-\bar w_i}{r_\ell-w_i}\right)
 - \Li_2\left(\frac{r_k-w_i}{r_k-\bar w_i}\frac{r_\ell\bar w_i+1}{r_\ell w_i+1}\right)
 \nonumber\\ &
 +\Li_2\left(\frac{r_k w_i+1}{r_k \bar w_i+1}\frac{r_\ell\bar w_i+1}{r_\ell w_i+1}\right)
 -\Li_2\left(\frac{r_k w_i+1}{r_k \bar w_i+1}\frac{r_\ell-\bar w_i}{r_\ell-w_i}\right)\Bigg]~.
\end{align}
The functions $\cA_\pm$ as defined in (\ref{eq:Apm-Z2}) as well as the regularity conditions (\ref{eq:O7-reg}) only involve the data on the right quadrant of the upper half plane.

The free parameters in $\cA_\pm$ are $3L_0$ in $r_\ell$ and $Z_\pm^\ell$ (with $\bar Z_+^\ell=-Z_-^\ell$),
and $4I_0$ in the positions, branch cut orientations, and monodromies associated with the branch points.
After fixing the fixed point of the $\ZZ_2$ symmetry to $w=i$, one $SL(2,\RR)$ redundancy remains.
The sum over the conditions in (\ref{eq:O7-reg-2}) vanishes, leaving $L_0+I_0-1$ independent conditions.
The result is a total of
\begin{align}
 2L_0+3I_0
\end{align}
free real parameters for a $\ZZ_2$ symmetric solution with $2L_0$ poles and $2I_0$ punctures.
They can be taken as the residues (5-brane charges) at half of the poles and the D7-brane charges, branch cut orientations and positions on a curve in $\Sigma$ for half the punctures.

\subsection{Solutions with O7 plane}\label{sec:sol}

\begin{figure}
\begin{tikzpicture}[scale=0.9]
\shade [ top color=blue! 1, bottom color=blue! 35] (5,1.5)  rectangle (9.8,5);
\shade [ top color=blue! 1, bottom color=blue! 10] (0.2,1.5)  rectangle (5,5);
\node at (9.5,4.5) {\large $\mathbf{\Sigma}$};

\draw[dashed,very thick,black] (5.0,3) -- (5.0,5.0);
\draw[black] (5.0,3.0) node {$\bullet$};
\draw (5.4,3) node {$w_1$};
\draw[black!60] (4.4,3) node {$w_{1+I_0}$};
\draw[dashed,very thick,black] (5.0,3) -- (5.0,1.5);

\draw[black!60] (2.5,3.5) node {$\bullet$};
\draw[black!60] (2.7,3.15) node {$w_{i+I_0}$};
\draw[black!60,dashed,thick,black,domain=0:1,smooth,variable=\c] plot ({2.5-3.4641*\c/(\c*\c+\c+1)},{3.5-4*\c*(\c+0.5)/(\c*\c+\c+1)});

\draw[black] (7.5,3.5) node {$\bullet$};
\draw (7.5,3.15) node {$w_{i}$};
\draw[dashed,very thick,black,domain=0:1,smooth,variable=\c] plot ({7.5+3.4641*\c/(\c*\c+\c+1)},{3.5-4*\c*(\c+0.5)/(\c*\c+\c+1)});

\draw [thick] (0,1.5) -- (10,1.5);

\draw[thick,black!60] (4.2,1.35) -- (4.2,1.65);
\draw[black!60] (4.2,1.15) node{$r_{L_0+1} $};
\draw[black!60] (3.35,1.15) node {$\ldots$};
\draw[black!60,thick] (2.5,1.35) -- (2.5,1.65);
\draw[black!60] (2.5,1.15) node{$r_{L_0+\ell}$};
\draw[black!60] (1.55,1.15) node {$\ldots$};
\draw[black!60,thick] (0.6,1.35) -- (0.6,1.65);
\draw[black!60] (0.6,1.15) node{$r_{2L_0}$};

\draw[thick] (5.8,1.35) -- (5.8,1.65);
\draw (5.8,1.15) node{$r_{L_0} $};
\draw (6.65,1.15) node {$\ldots$};
\draw[thick] (7.5,1.35) -- (7.5,1.65);
\draw (7.5,1.15) node{$r_\ell $};
\draw (8.45,1.15) node {$\ldots$};
\draw[thick] (9.4,1.35) -- (9.4,1.65);
\draw (9.4,1.15) node{$r_1 $};

\end{tikzpicture}

 \caption{Upper half plane with symmetric arrangement of poles and punctures and two merged punctures at $w=i$. The branch cuts extending in opposite directions along the imaginary axis are mapped into each other by the $\ZZ_2$ symmetry (\ref{eq:fC}).
 The quotient solution can be taken as the right quadrant.
 \label{fig:O7-sol}}
\end{figure}
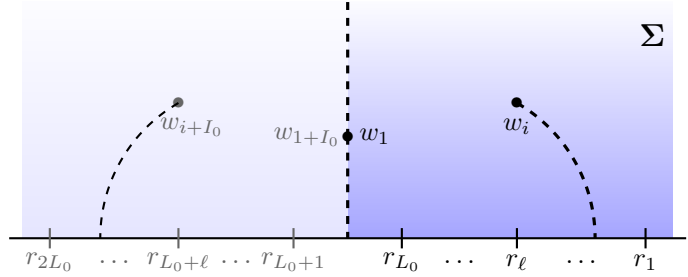

Solutions with a puncture corresponding to a combination of an O7-plane and D7-branes are constructed from the $\ZZ_2$-symmetric solutions by taking the limit where one pair of symmetry-related D7-brane punctures becomes coincident at $w=i$, say $w_1\rightarrow i$, and by subsequently taking a $\ZZ_2$ quotient.
The residual regularity conditions discussed in the previous sections were derived assuming isolated branch points,
but since the configuration with coincident branch points can be seen as limiting case of a family of solutions with separated branch points, sufficient regularity conditions can be obtained by the same limiting procedure.

The orientation of the branch cuts associated with the coincident punctures at $w=i$ can be fixed without loss of generality: two $SL(2,\RR)$ degrees of freedom were used to realize the fixed point of the $\ZZ_2$ symmetry at $w=i$, and the remaining one can be used to fix the orientation of the branch cuts starting at $w=i$ along the imaginary axis.
In summary, a solution with $-T^{n_1^2}$ puncture, and possibly additional separate D7-brane punctures, is realized by the $\cA_\pm$ in (\ref{eq:Apm-Z2}) with 
\begin{align}\label{eq:O7-punc}
 w_1&=i~, & \gamma_1&=1~,
\end{align}
subject to the regularity conditions in (\ref{eq:O7-reg}).
Note that $w_1=i$ solves the $i=1$ regularity condition in (\ref{eq:O7-reg-1}).
The quotient by the geometric action of $\fC$ in (\ref{eq:fC}) then reduces the solution to the right quadrant of the upper half plane,
as shown in fig.~\ref{fig:O7-sol}. The two branch cuts emanating from the puncture at $w=i$ are identified; in the fundamental domain there is one branch cut associated with the puncture at $w=i$, with $-T^{n_1^2}$ monodromy.

The remaining parameters in $\cA_\pm$ are $3L_0+1+4(I_0-1)$.
The remaining conditions are those in (\ref{eq:O7-reg-2}), which sum to zero, and those in (\ref{eq:O7-reg-1}) for $i>1$, a total of $L_0+I_0-2$.
In realizing the O7 puncture at $w_1=i$ with the branch cut along the imaginary axis, all $SL(2,\RR)$ redundancies were used. 
The number of free parameters consequently is 
\begin{align}\label{eq:param-O7}
2L_0+2+3(I_0-1)~.
\end{align}
They correspond to $L_0$ $(p,q)$ 5-brane charges, $2$ parameters associated with the O7 puncture, and $3$ parameters for each additional D7 puncture (those were discussed in \cite{DHoker:2017zwj}).
One of the parameters associated with the O7 puncture is its D7-brane charge, $n_1^2$.
The other is related to the orientation of the branch cut: 
The branch cut was fixed along the imaginary axis, but its position is only meaningful relative to the poles.
The remaining parameter can be taken as the position of one of the poles, or equivalently as the orientation of the branch cut relative to the poles.

The expressions for $\cA_\pm$ and the regularity conditions simplify for solutions with one puncture corresponding to a combination of an O7-plane with D7-branes and no additional separate D7-brane punctures. This case will be spelled out explicitly.
It will be convenient to map the upper half plane to the unit disc with coordinate $z$ and poles labeled by $t_\ell$ via
\begin{align}\label{eq:v-coord-def}
 z&=\frac{i-w}{i+w}~,& t_\ell&=\frac{i-r_\ell}{i+r_\ell}~.
\end{align}
The branch point is mapped to the origin and the branch cuts extend along the positive and negative real axis.
The right quadrant of the upper half plane is mapped to the upper hemisphere in the $z$ coordinate.
This realizes the solution as in fig.~\ref{fig:strategy-b}.
The functions $\cA_\pm$ are given by
\begin{align}\label{eq:cApm-O7-only}
\cA_\pm&= \sum_{\ell=1}^{L_0}\left[\left(Z_\pm^\ell+ \frac{n_1^2}{4\pi} Y^\ell\ln \left(-z^2\right)\right)\ln\left(\frac{t_\ell-z}{t_\ell+z}\right)+ \frac{n_1^2}{2\pi} Y^\ell
\left(
\Li_2\left(\frac{z}{t_\ell}\right) -\Li_2\left(-\frac{z}{t_\ell}\right)
\right)\right]\,.
\end{align}
The remaining regularity conditions are
\begin{align}\label{eq:reg-O7-only}
  0&=\sum_{\stackrel{\ell=1}{\ell\neq k}}^{L_0}\left[ Z^{[\ell,k]}\ln\left|\frac{t_\ell-t_k}{t_k+t_\ell}\right|^2
 +Y^k Y^\ell \sum_{i=1}^{I_0}\frac{n_i^2}{2\pi} \left(\Li_2\left(\frac{t_k}{t_\ell}\right)-\Li_2\left(-\frac{t_k}{t_\ell}\right)-\rm{c.c.}\right)\right]~, 
\end{align}
for $k=1,\ldots L_0$. 
The solution in the fundamental domain with coordinate $u$ on the unit disc is obtained via the coordinate transformation (\ref{eq:z-u}),
\begin{align}\label{eq:cApm-O7-only-u}
\cA_\pm&= \sum_{\ell=1}^{L_0}\left[\left(Z_\pm^\ell+ \frac{n_1^2}{4\pi} Y^\ell\ln (-u)\right)\ln\left(\frac{t_\ell-i\sqrt{-u}}{t_\ell+i\sqrt{-u}}\right)+ \frac{n_1^2}{2\pi} Y^\ell
\left(
\Li_2\left(\frac{i\sqrt{-u}}{t_\ell}\right) -\Li_2\left(-\frac{i\sqrt{-u}}{t_\ell}\right)
\right)\right].
\end{align}
The $\cA_\pm$ have one branch cut in the interior of the disc in the $u$ coordinate, extending from the origin along the positive real axis.

\subsection{Behavior at puncture}\label{sec:near-puncture}

To verify that the Type IIB supergravity fields have the desired monodromy as the puncture is encircled it is sufficient to establish the correct transformation of $\cA_\pm$, since the $SL(2,\RR)$ transformations of the supergravity fields are induced by the $SU(1,1)$ transformations of $\cA_\pm$ in (\ref{eq:cApm-SU11}).
This will be discussed for the expressions in (\ref{eq:cApm-O7-only-u}); since the transformation only depends on the local behavior near the puncture the results extend to solutions with additional D7 punctures.
Encircling the puncture, starting with $u$ slightly above the positive real axis and following a counter-clockwise contour to $ e^{2\pi i}u$ has the effect that $\sqrt{-u}\rightarrow -\sqrt{-u}$ and $\ln(-u)\rightarrow \ln(-u)+2\pi i$ in the expression for $\cA_\pm$ in (\ref{eq:cApm-O7-only-u}).
With these replacements,
\begin{align}\label{cApm-transf}
 \cA_\pm& \rightarrow \cA_\pm^\prime =  -\left(\cA_\pm + \frac{in_1^2}{2}\left(\cA_+-\cA_-\right)\right)~.
\end{align}
Comparing this to (\ref{eq:cApm-SU11}) and the translation to $SL(2,\RR)$ below (\ref{eq:cApm-SU11}) shows that it indeed amounts to the expected $SL(2,\RR)$ transformation $-T^{n_1^2}$.
The behavior of the supergravity fields is shown for an example in fig.~\ref{fig:vert-plot}.

The construction of solutions with O7-plane as quotient of solutions with D7-branes realizes one O7/D7 puncture (and possibly additional separated D7 punctures).
Solutions with multiple O7-planes may be realized by extracting the behavior of $\cA_\pm$ near puncture with $-T^k$ monodromy and incorporating it into a more general ansatz.
The relevant information is the behavior of the differentials $\partial_u\cA_\pm$ near the puncture.
For small $u$ they take the general form
\begin{align}
  \partial_u\cA_\pm&=\frac{1}{\sqrt{-u}}\left[\partial_u\cA^{(0)}_\pm+\frac{n_1^2}{4\pi}\ln(- u)\left(\partial_u\cA^{(0)}_+-\partial_u\cA^{(0)}_-\right)\right]~,
\end{align}
with $\partial_u\cA^{(0)}_\pm$ finite and single-valued at $u=0$. 
Compared to the logarithmic divergence near D7-brane punctures, there is an additional divergent overall square root factor.
More general solutions may be constructed  by starting with a general ansatz incoorporating such singularities in the differentials and carrying out a regularity analysis along the lines of \cite{DHoker:2017zwj}.

\section{5d SCFTs and holographic duals}\label{sec:5dSCFTs}

The supergravity solutions constructed in sec.~\ref{sec:AdS6-O7-sol} are naturally identified with the conformal limit of $(p,q)$ 5-brane webs on combinations of O7-planes and D7-branes.
The poles in the supergravity solutions represent the external 5-branes of the brane web, with the charges encoded in the residues.
The puncture with $-T^k$ monodromy corresponds to a combination of O7-planes and D7-branes leading to that monodromy in the central face of the web.
Additional punctures with $T^n$ monodromy correspond to D7-branes in other faces of the web, with the location of the face encoded in the position of the puncture on the disc.

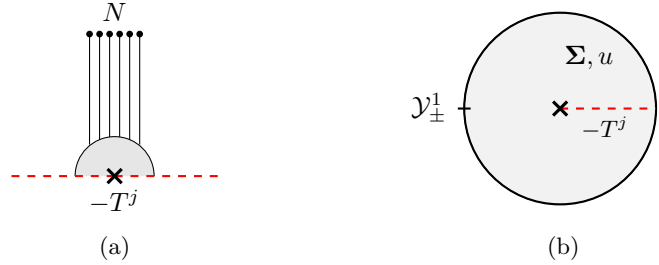
\begin{figure}
\subfigure[][]{
 \begin{tikzpicture}[scale=1.05]\label{fig:vert-1}
    \foreach \i in {-5/2,-3/2,-1/2,1/2,3/2,5/2}{
      \draw (0.127*\i,0.4) -- (0.127*\i,1.8) [fill=black] circle (1pt);
    }
    \draw[fill=gray!20] (0.5,0) arc (0:180:0.5);
    \draw[thick,dashed,red] (0,0) -- (1.4,0);
    \draw[thick,dashed,red] (0,0) -- (-1.4,0);
    \draw[very thick] (-0.09,0.09) -- (0.09,-0.09);
    \draw[very thick] (-0.09,-0.09) -- (0.09,0.09);

    \node[anchor=north] at (0,-0.05) {$-T^j$};
    
    \node [anchor=south] at (0,1.85) {$N$};
 \end{tikzpicture}
}\qquad\qquad\qquad
\subfigure[][]{\label{fig:fund-min}
\begin{tikzpicture}[scale=0.85]
\draw[fill=lightgray!20] (0,0) circle (1.5);
\draw[thick] (0,0) circle (1.5);
\draw[thick] (-1.4,0) -- (-1.6,0);
\node at (0.45,0.8) {$\mathbf \Sigma, u$};
\node at (-2.05,0) {\small $\cY_\pm^1$};

\draw[thick,dashed, red] (0,0) -- (1.5,0);
\draw[very thick] (-0.11,0.11) -- (0.11,-0.11);
\draw[very thick] (-0.11,-0.11) -- (0.11,0.11);

\node at (0.7,-0.3) {\footnotesize $-T^j$};

\node at (0,-1.5) {\small \vphantom{$M$ NS5}};
\node at (2.15,0) {\small \hphantom{$Z_\pm^1$}};
\end{tikzpicture}
}
 \caption{
 On the left the \vertx$_{N,j}^\pm$ junction.
 The representation with two copies of the branch cut will be convenient for discussing gauge theory deformations.
 For \vertx$_{N,j}^-$ the monodromy is due to an O7$^-$ with $j+4$ D7 branes, for \vertx$_{N,j}^+$ with $j\geq 4$ it is due to an O7$^+$ with $j-4$ D7 branes.  On the right the schematic form of the associated supergravity solution with one puncture at the center and one 5-brane pole on the boundary.\label{fig:vert}}
\end{figure}

In this section a sample of 5-brane junctions on O7/D7 combinations is discussed, along with the 5d SCFTs that they engineer and their gauge theory deformations. The supergravity solutions associated with the junctions are spelled out explicitly, which allows to holographically compute the $S^5$ free energy of the 5d SCFTs.
The gauge theory deformations of the 5d SCFTs engineered by the junctions are generalizations of the gauge theories discussed in \cite{Bergman:2015dpa}. The gauge theories engineered by D5-branes on O7-planes can be summarized as follows:
\begin{align}\label{tab:gauge-groups}
\begin{tabular}{c@{\hskip .1in}|@{\hskip .1in}c@{\hskip .2in}c}
\hline
  & O7$^-$ & O7$^+$\\
  \hline
  $2N$ D5 & $USp(2N)$ & $SO(2N)$\\
  $M$ D5 + $1$ NS5 & $SU(M)+[1]^{}_A$ & $SU(M)+[1]^{}_S$\\
\hline
\end{tabular}
\end{align}
The 5-brane counts here include branes and their mirrors.
Adding a fractional D5 on an O7$^+$ with no fractional NS5 engineers an $SO(2N+1)$ gauge group.
If a fractional NS5 ends on the O7, the gauge theory is $SU(M)$ with an antisymmetric/symmetric hypermultiplet for O7$^-$/O7$^+$.
For $M$ odd there also is a fractional D5.
The 5-brane junctions and associated supergravity solutions to be discussed in the following are a minimal one in fig.~\ref{fig:vert} and a simple generalization shown in fig.~\ref{fig:perp}.
Deformations leading to gauge theories are shown in fig.~\ref{fig:vert-web} and \ref{fig:perp-webs}.
Solutions for more general 5-brane webs on O7-planes can be constructed along the same lines.

\subsection{\texorpdfstring{\vertx$_{N,j}^\pm$}{I[N,j]} theories}

The first example is a stack of $N$ NS5 branes on a 7-brane with $-T^j$ monodromy and $j\geq 0$, shown in fig.~\ref{fig:vert-1}.
Alluding to the shape, the brane configuration and the associated 5d SCFTs will be referred to as $\vertx_{N,j}^\pm$.
For $\vertx_{N,j}^-$ the monodromy is realized by a combination of an O7$^-$  with $j+4$ D7-branes, for $\vertx_{N,j}^+$, with $j\geq 4$, it is realized by an O7$^+$ with $j-4$ D7 branes.

The form of the gauge theory deformations depends on whether $N$ is even or odd.
An example for a mass deformation with $j=4$ and $N$ even/odd is shown in fig.~\ref{fig:vert-web-1}/fig.~\ref{fig:vert-web-2}.
The central node is read off using (\ref{tab:gauge-groups}).
The nodes away from the orientifold remain $SU(\cdot)$ nodes, but the nodes to the right are identified with the nodes to the left.
This is analogous to similar constructions in other dimensions, for example for 6d theories in \cite{Brunner:1997gf}.

\begin{figure}
\subfigure[][]{\label{fig:vert-web-1}
 \begin{tikzpicture}[scale=0.9]
  \draw[dashed,red,thick] (0,0) -- (4,0);
  \draw[dashed,red,thick] (0,0) -- (-4,0);
  \draw[very thick] (-0.11,0.11) -- (0.11,-0.11);
  \draw[very thick] (-0.11,-0.11) -- (0.11,0.11);

  \draw (-1.4,0) -- (-0.9,0.25) -- (0.9,0.25) -- (1.4,0);
  \draw (-0.9,0.25) -- (-0.9,1.5) node [anchor=south] {\scriptsize $(0,1)$};
  \draw (0.9,0.25) -- (0.9,1.5) node [anchor=south] {\scriptsize $(0,1)$};
  \node at (0,0.45) {\scriptsize $(2,0)$};
    
  \draw (-3.6,0) --(-2.1,0.75) -- (2.1,0.75) -- (3.6,0);
  \node at (0,0.95) {\scriptsize $(2K\,{-}\,2,0)$};
  \draw (-2.1,0.75) -- (-2.1,1.5) node [anchor=south] {\scriptsize $(0,K\,{-}\,1)$};
  \draw (2.1,0.75) -- (2.1,1.5) node [anchor=south] {\scriptsize $(0,K\,{-}\,1)$};
  
  \node [anchor=north] at (0,-0.05) {O7$^+$\,/\,O7$^-\,{+}\,8$\,D7};
 \end{tikzpicture}
 }
\qquad
\subfigure[][]{\label{fig:vert-web-2}
 \begin{tikzpicture}[scale=0.9]
  \draw[dashed,red,thick] (0,0) -- (4,0);
  \draw[dashed,red,thick] (0,0) -- (-4,0);
  \draw[very thick] (-0.11,0.11) -- (0.11,-0.11);
  \draw[very thick] (-0.11,-0.11) -- (0.11,0.11);
  
  \draw (-1.4,0) -- (-0.9,0.25) -- (0.9,0.25) -- (1.4,0);
  \draw (-0.9,0.25) -- (-0.9,1.5) node [anchor=south] {\scriptsize $(0,1)$};
  \draw (0.9,0.25) -- (0.9,1.5) node [anchor=south] {\scriptsize $(0,1)$};
    
  \draw (-3.6,0) --(-2.1,0.75) -- (2.1,0.75) -- (3.6,0);
  \draw (-2.1,0.75) -- (-2.1,1.5) node [anchor=south] {\scriptsize $(0,K\,{-}\,1)$};
  \draw (2.1,0.75) -- (2.1,1.5) node [anchor=south] {\scriptsize $(0,K\,{-}\,1)$};
  
  \draw (0,0) -- (0,1.5) node [anchor=south] {\scriptsize $(0,1)$};;
  
  \node [anchor=north] at (0,-0.05) {O7$^+$\,/\,O7$^-\,{+}\,8$\,D7};
 \end{tikzpicture}
 }

 \caption{Gauge theory deformations of \vertx$_{N,j}^\pm$ for $j=4$ and $N=2K$ even on the left, and with a fractional NS5, $N=2K+1$ odd, on the right. After decoupling the quiver tails by moving the $(0,K-1)$ branes to infinity, the remaining central nodes correspond to gauge theories discussed in \cite{Bergman:2015dpa}.
 For $N$ even and an O7$^+$, it is $SO(2N)$ with $2N-4$ hypermultiplets in the vector representation, for an O7$^-$ with $8$ D7 it is $USp(2N)$ with $2N+4$ fundamental hypermultiplets.
 For $N$ odd and an O7$^+$ the central node is $SU(2N-2)$ with one symmetric and $2N-6$ fundamental hypermultiplets,
 for an O7$^-$ with $8$ D7 it is $SU(2N-2)$ with one antisymmetric and $2N+2$ fundamental hypermultiplets.
 Leaving the $(0,K-1)$ branes in the picture and separating them horizontally gives the full quiver gauge theories.
 \label{fig:vert-web}}
\end{figure}

For $N=2K$ even the gauge theory deformations of the \vertx$_{N,j}^\pm$ theories are,
assuming that $jK$ is even for $\vertx_{N,j}^-$ and $j\geq 4$ for $\vertx_{N,j}^+$,
\begin{align}
 \vertx_{2K,j}^-:& &
 [j+4] - USp(jK) &- SU(jK-j)  - \ldots - SU(2j) - SU(j)~,
 \nonumber\\
  \vertx_{2K,j}^+:& &
 [j-4] - \; SO(jK) \;  &- SU(jK-j) - \ldots - SU(2j) -  SU(j)~.
 \label{eq:I-USp-SO-quivers}
\end{align}
Along the $SU(\cdot)$ quiver tails the rank decreases in steps of $j$.
All Chern-Simons levels for the $SU(\cdot)$ nodes are zero.
The global symmetry manifest in the gauge theory deformation is $SO(2j+8)\times U(1)_I^K\times U(1)_B^{K-1}$ for $\vertx_{2K,j}^-$
and $USp(2j-8)\times U(1)_I^K\times U(1)_B^{K-1}$ for $\vertx_{2K,j}^+$.
All nodes saturate the flavor bounds of \cite{Intriligator:1997pq}.
For $K=1$ the \vertx$_{2,j}^-$ theory reduces to $USp(j)-[j+4]$, which for $j=2$ is the $E_7$ theory. 
The \vertx$_{2K,j}^+$ theory with $K=1$ and $j\geq 4$ becomes $SO(j)-[j-4]$.

For $N=2K+1$ odd, with a fractional NS5 on the O7$^\pm$, the gauge theory deformations are
\begin{align}
 \vertx_{2K+1,j}^-:& &
 [j+4] - SU&(jK) - SU(jK-j) - \ldots - SU(2j) - SU(j)
 \nonumber\\
 &&&\,\vert\nonumber\\
 &&&\hspace*{-1mm}[1]^{}_A\nonumber\\[2mm]
  \vertx_{2K+1,j}^+:& &
 [j-4] - SU&(jK) - SU(jK-j) -  \ldots - SU(2j) - SU(j)
  \nonumber\\
 &&&\,\vert\nonumber\\
 &&&\hspace*{-1mm}[1]^{}_S
 \label{eq:I-SU-A-S-quivers}
\end{align}
All Chern-Simons levels are zero. 
The global symmetry manifest in the gauge theory deformations for $\vertx_{2K+1,j}^-$ is $SU(j+4)\times U(1)_A\times U(1)_I^K\times U(1)_B^{K}$.
For $\vertx_{2K+1,j}^+$ with $j>4$ it is $SU(j-4)\times U(1)_S\times U(1)_I^K\times U(1)_B^{K}$, which reduces to $U(1)_S\times U(1)_I^K\times U(1)_B^{K-1}$ for $j=4$.
For $K=1$ the gauge theory for \vertx$_{2,j}^-$ is $[1]^{}_A-SU(j)-[j+4]$.
For $j=2$ this is the $E_7$ theory and a singlet.
The gauge theory for \vertx$_{2,j}^+$ with $K=1$ and $j\geq 4$ is $[1]^{}_S-SU(j)-[j-4]$.
The flavor bound of \cite{Intriligator:1997pq} for $SU(N_c)$ with an antisymmetric is $N_F+2|c_{\rm cl}|\leq 8-N_c$.
The node with antisymmetric hypermultiplet in (\ref{eq:I-SU-A-S-quivers}) can have arbitrarily large rank, and this bound is in general violated.
The node is compatible with the bound $N_F+2|c_{\rm cl}|\leq N_c+5$ of \cite{Bergman:2015dpa}.
Analogous statements apply for the theories with symmetric hypermultiplet, which were ruled out in~\cite{Intriligator:1997pq}.

\subsection{\texorpdfstring{$\vertx_{N,j}^\pm$}{I[N,j]} solutions}\label{sec:min-sol}

The supergravity solutions associated with the brane junction in fig.~\ref{fig:vert-1} are minimal in the sense that they only involve one O7/D7 puncture and one 5-brane pole. They are identical for $\vertx_{N,j}^\pm$.
The functions $\cA_\pm$ are given by (\ref{eq:cApm-O7-only-u}) with $L_0=I_0=1$.
The regularity condition in (\ref{eq:reg-O7-only}) is trivial and satisfied automatically.
The position of the pole is a free parameter, encoding the orientation of the branch cut associated with the O7 puncture relative to the 5-brane pole, as discussed after (\ref{eq:param-O7}). As discussed in \cite{Gutperle:2018vdd}, it matters between which two poles a branch cut crosses the boundary $\partial\Sigma$, otherwise the precise orientation is irrelevant. 
We set $t_1=i$ such that the pole is at $u=-1$.
This realizes the solution as in fig.~\ref{fig:fund-min}.
The functions $\cA_\pm$ are given by
\begin{align}\label{eq:cApm-vert}
 \cA_\pm&=
 \left(Z_\pm^1+ \frac{n_1^2}{4\pi} Y^1\ln (-u)\right)\ln\left(\frac{1-\sqrt{-u}}{1+\sqrt{-u}}\right)+ \frac{n_1^2}{2\pi} Y^1
\left(
\Li_2\left(\sqrt{-u}\right) -\Li_2\left(-\sqrt{-u}\right)
\right)~.
\end{align}
The residues at the pole at $u=-1$ are given by $\cY_\pm^1=Z_\pm^1$.
The physical parameters are the residues at the pole, encoding the 5-brane charge via (\ref{eq:cY-pq}), and the D7 charge $n_1^2$ at the puncture. This matches the parameters of the corresponding 5-brane junctions.
A non-degenerate solution needs $Z_+^1\neq Z_-^1$, since $\kappa^2$ vanishes identically otherwise. 
That means the residues encode non-trivial NS5 charge.
For the same reason, the solution degenerates for $n_1^2\rightarrow 0$.
This behavior matches the string theory picture, where a 5d SCFT can only be engineered with one stack of 5-branes if their NS5 charge is non-zero and the puncture they probe has $n_1^2>0$.

\begin{figure}
\centering
\begin{tabular}{c@{\hskip 6mm}c@{\hskip 6mm}c}
\includegraphics[width=0.3\linewidth]{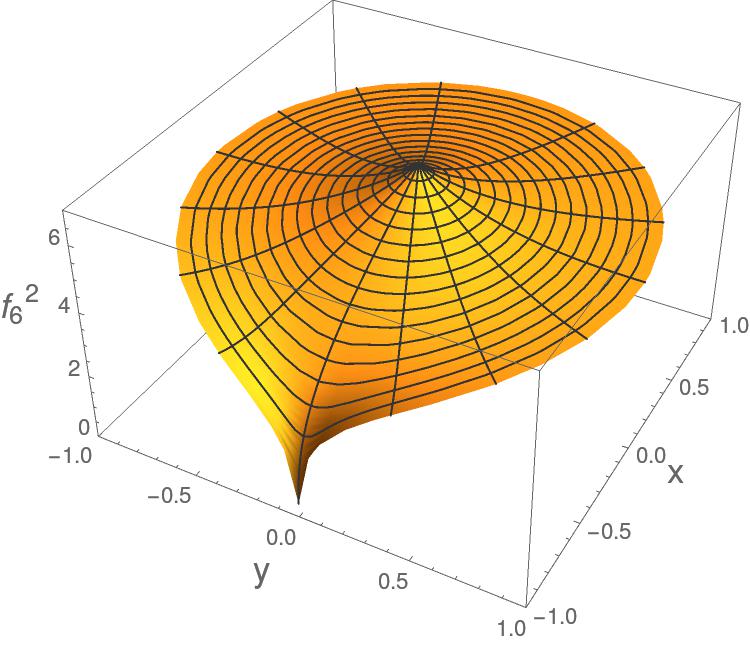} &
\includegraphics[width=0.3\linewidth]{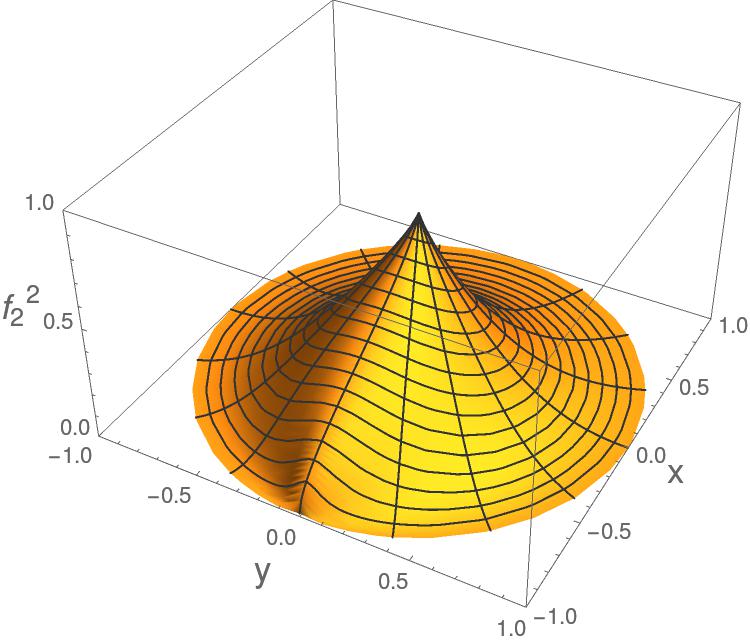}&
\includegraphics[width=0.3\linewidth]{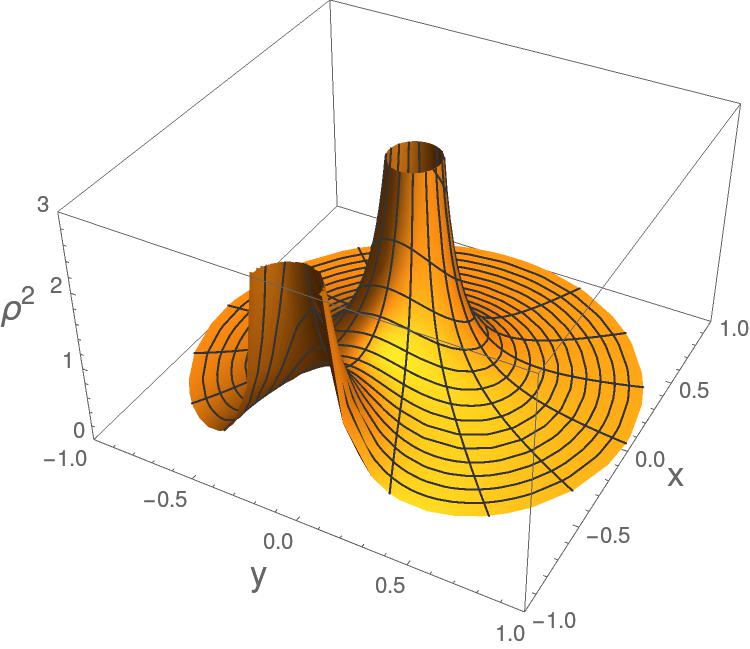}
\\[2mm]
\includegraphics[width=0.3\linewidth]{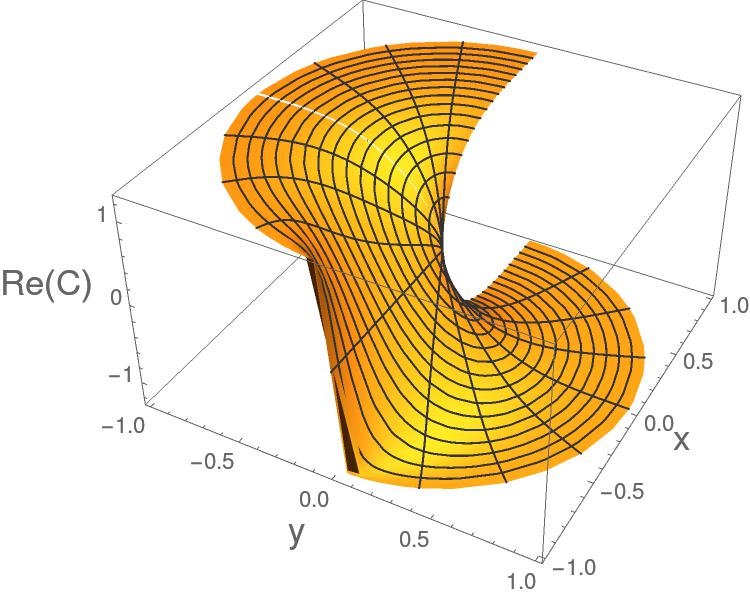}&
\includegraphics[width=0.3\linewidth]{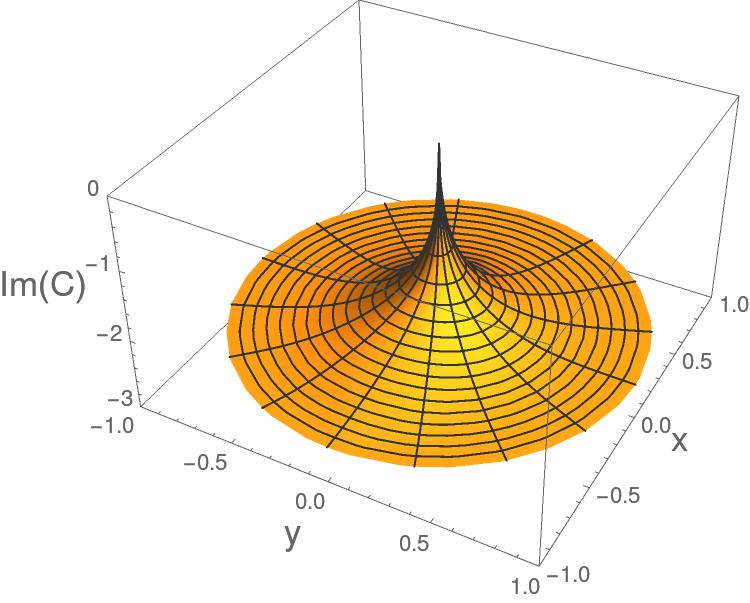}
\\[2mm]
\includegraphics[width=0.3\linewidth]{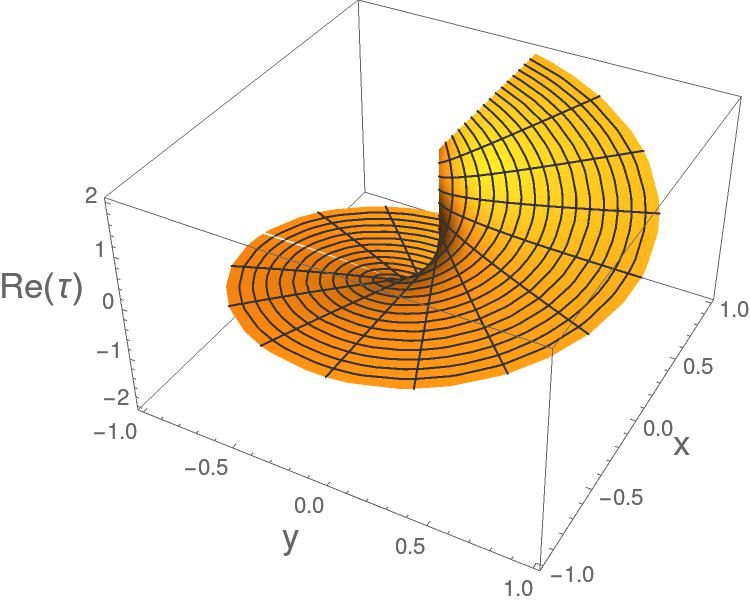}&
\includegraphics[width=0.3\linewidth]{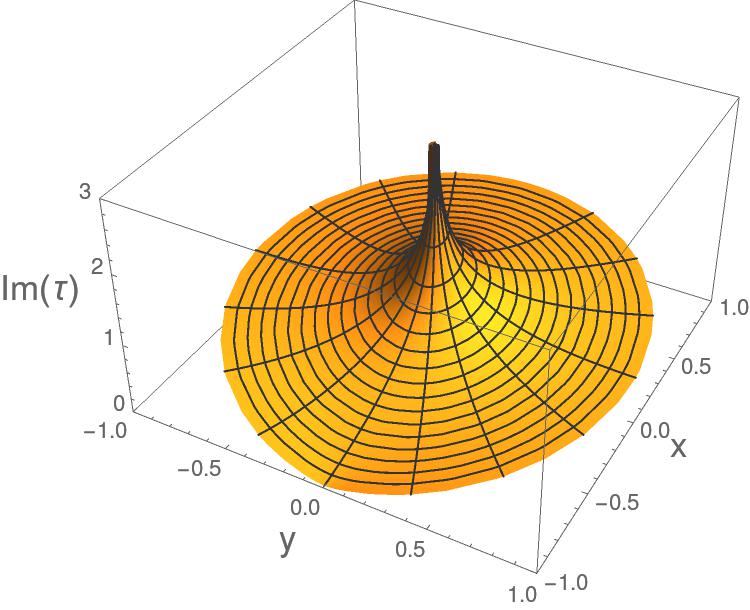}
\end{tabular}

\caption{
 Plots of the supergravity fields for the \vertx$_{N,j}^\pm$ solution, with one pole corresponding to $N$ NS5-branes and a puncture with $-T^j$ monodromy, for $j=4$ and $N=5$.
 The metric and complex 2-form in (\ref{eqn:ansatz}) are parametrized as $ds^2=f_6^2 ds^2_{AdS_6}+f_2^2 ds^2_{S^2}+4\rho^2 |du|^2$ and $C_{(2)}=\cC \vol_{S^2}$, with $u=x-iy$.
 The plots in the first row show the metric functions,
 the plots in the second row show the real and imaginary part of $\cC$, and the last two show the real and imaginary parts of $\tau$.
 The metric functions are single-valued and positive in the interior of the disc. 
 At the boundary the $S^2$ collapses, $f_2^2\rightarrow 0$, closing off the internal space smoothly.
 At the pole the behavior is as expected for a 5-brane (as discussed in \cite{DHoker:2017mds}).
 At the puncture, $f_2^2$ and $f_6^2$ are finite with a cusp, $\rho^2$ behaves as $|u|^{-1}\ln |u|$.
 $\cC$ and $\tau$ transform as desired across the branch cut.
 The imaginary part of $\cC$ is finite with a cusp at the puncture, the imaginary part of $\tau$ diverges.
 \label{fig:vert-plot}}

\end{figure}

The supergravity description for the \vertx$_{N,j}^\pm$ theories is obtained for 
\begin{align}
 \cY_\pm^1=Z_\pm^1&=\pm\frac{3}{4}\alpha^\prime N~, & n_1^2&=j~.
\end{align}
All other supergravity solutions with one 5-brane pole and one O7/D7 puncture can be obtained from an \vertx$^\pm_{N,j}$ solution by a global $SL(2,\RR)$ transformation which leaves the monodromy of the puncture invariant.\footnote{%
For O7$^-$ there are two distinct variants, corresponding to different resolutions into $[p,q]$ 7-branes and related by $T$ \cite{Bergman:2015dpa}. 
But at the level of the supergravity solution the $T^{-j}$ monodromy is invariant under conjugation by $T$.}
Plots of the solution are shown in fig.~\ref{fig:vert-plot}.
They show that the supergravity fields have the expected monodromy at the puncture and the required behavior at the pole,
and that they otherwise satisfy the general regularity conditions.

With the supergravity solutions as holographic duals in hand, the $S^5$ free energy for the \vertx$_{N,j}^\pm$ theories can be computed. 
As derived in \cite{Gutperle:2017tjo,Fluder:2018chf}, the holographic free energy is given by
\begin{align}\label{eq:FS5-sugra-gen}
 F_{S^5}&=-\frac{32\pi^3}{9G_N}\int_\Sigma d^2u\,|\partial_u\cG|^2~,
 &
 \partial_u \cG&=(\bar\cA_+-\cA_-)\partial_u\cA_+ + (\cA_+-\bar \cA_-)\partial_u\cA_-~,
\end{align}
with $16\pi G_N=(2\pi)^7(\alpha^\prime)^4$.
The integrand depends on $N$ and $j$ through an overall factor $N^4j^2$, leaving only a numerical coefficient to be evaluated.
The result is
\begin{align}\label{eq:F-vert-sugra}
 F^{\vertx^\pm_{N,j}}_{S^5}&=
 -\frac{837}{128 \pi ^4}\zeta (5) j^2 N^4 ~.
\end{align}
This is the holographic result for the free energy of the UV fixed point SCFTs of all four quiver theories in (\ref{eq:I-USp-SO-quivers}), (\ref{eq:I-SU-A-S-quivers}), in the limit where $N$ is large and $j$ of order one.
It exhibits the quartic scaling characteristic of the 5d SCFTs with holographic duals in Type IIB and follows the pattern that the results for solutions with punctures involve polylogarithms of degree up to 5 \cite{Uhlemann:2019ypp}.

\subsection{\texorpdfstring{\perpx$^\pm_{N,M,j}$}{perp[N,M,j]} theories}

\begin{figure}
 \subfigure[][]{
   \begin{tikzpicture}[scale=1.05]\label{fig:Op-1}
    \foreach \i in {-3/2,-1/2,1/2,3/2}{
        \draw (0.14*\i,0) -- (0.14*\i,1.6) [fill=black] circle (1pt);
    }
    \foreach \i in {1/2,3/2,5/2}{
      \draw (0,0.14*\i) -- (-1.6,0.14*\i) [fill=black] circle (1pt) ;
      \draw (0,0.14*\i) -- (1.6,0.14*\i) [fill=black] circle (1pt) ;
    }
    \draw[fill=gray!20] (0.5,0) arc (0:180:0.5);
    
    \draw[thick,dashed,red] (0,0) -- (2.6,0);
    \draw[thick,dashed,red] (0,0) -- (-2.6,0);
    \node[anchor=north] at (0,-0.1) {$-T^j$};
    \draw[very thick] (-0.09,0.09) -- (0.09,-0.09);
    \draw[very thick] (-0.09,-0.09) -- (0.09,0.09);
    
    \node  at (2.15,0.2) { $\lfloor N/2\rfloor$};
    \node  at (-2.15,0.2) { $\lfloor N/2\rfloor$};
    \node [anchor=south] at (0,1.7) {$M$};
 \end{tikzpicture}
 }\hskip 15mm
 \subfigure[][]{\label{fig:fund-perp}
\begin{tikzpicture}[scale=0.85]
\draw[fill=lightgray!20] (0,0) circle (1.5);
\draw[thick] (0,0) circle (1.5);
\draw[very thick] (-1.4,0) -- (-1.6,0);
\node at (0.45,0.8) {$\mathbf \Sigma, u$};
\node at (-2,0) {\small $\cY_\pm^2$};
\draw[very thick] (1.4,0) -- (1.6,0);
\node at (2,0) {\small $\cY_\pm^1$};

\draw[thick,dashed, red] (0,0) -- (1.5,0);
\draw[very thick] (-0.11,0.11) -- (0.11,-0.11);
\draw[very thick] (-0.11,-0.11) -- (0.11,0.11);

\node at (0.7,-0.3) {\footnotesize $-T^j$};

\node at (0,-1.5) {\small \vphantom{$M$ NS5}};
\end{tikzpicture}
}
 \caption{On the left the 5-brane junction for the \perpx$^\pm_{N,M,j}$ theories; if $N$ is odd there is an unpaired D5 stuck on the orientifold, if $M$ is odd there is a fractional NS5 on the orientifold.
 On the right the associated \perpx$^\pm_{N,M,j}$ solution with two 5-brane poles and one puncture with $-T^j$ monodromy, with the branch cut intersecting the pole corresponding to D5 branes.\label{fig:perp}}
\end{figure}
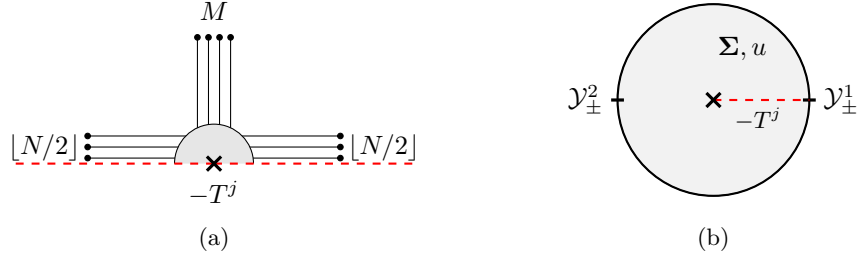

The \perpx$_{N,M,j}^\pm$ theories are realized by a junction of $N$ D5-branes and $M$ NS5-branes on a 7-brane with $-T^j$ monodromy, fig.~\ref{fig:Op-1}.
For \perpx$_{N,M,j}^-$ the monodromy is realized by a combination of an O7$^-$ with $j+4$ D7 branes,
for \perpx$^+_{N,M,j}$ with $j\geq 4$ it is realized by an O7$^+$ with $j-4$ D7-branes.
For $N=0$ and generic $M$ the \perpx$_{N,M,j}^\pm$ junctions reduce to \vertx$_{M,j}^\pm$.
For $M=2$ the \perpx$_{N,2,j}^\pm$ junctions are related to \vertx$_{2,N+j}$ by Hanany-Witten transitions.

\begin{figure}
 \begin{tikzpicture}[scale=0.9]
  \draw[dashed,red,thick] (0,0) -- (4,0);
  \draw[dashed,red,thick] (0,0) -- (-4,0);
  \draw[very thick] (-0.11,0.11) -- (0.11,-0.11);
  \draw[very thick] (-0.11,-0.11) -- (0.11,0.11);
  
  \draw (-1.4,0) -- (-0.9,0.25) -- (0.9,0.25) -- (1.4,0);
  \draw (-0.9,0.25) -- (-0.9,1.9) node [anchor=south] {\scriptsize $(0,1)$};
  \draw (0.9,0.25) -- (0.9,1.9) node [anchor=south] {\scriptsize $(0,1)$};
  \node at (0,0.45) {\scriptsize $(2,0)$};
    
  \draw (-3.6,0) --(-2.1,0.75) -- (2.1,0.75) -- (3.6,0);
  \node at (0,0.95) {\scriptsize $(2K\,{-}\,2,0)$};
  \draw (-2.1,0.75) -- (-2.1,1.9) node [anchor=south] {\scriptsize $(0,K\,{-}\,1)$};
  \draw (2.1,0.75) -- (2.1,1.9) node [anchor=south] {\scriptsize $(0,K\,{-}\,1)$};
  
  \draw (-4,1.3) -- (4,1.3);
  \node at (3.5,1.55) {\scriptsize $(N,0)$};
  
  \node [anchor=north] at (0,-0.05) {O7$^+$\,/\,O7$^-\,{+}\,8$\,D7};
 \end{tikzpicture}
 \caption{Gauge theory deformation of \perpx$_{N,M,j}^\pm$ for $j=4$ and $M=2K$ even.
 For $M$ odd there is an additional NS5 on the O7, as in fig.~\ref{fig:vert-web-2}.
 For $N$ odd there is an unpaired D5 on the branch cut.
 \label{fig:perp-webs}}
\end{figure}
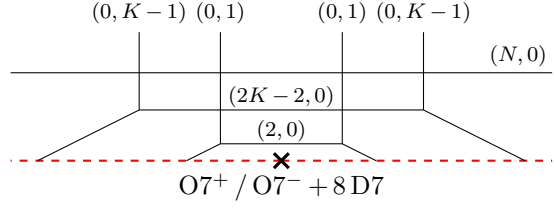

Gauge theory deformations are illustrated in fig.~\ref{fig:perp-webs}.
For $M=2K$ even, the quiver gauge theory deformations for \perpx$^-_{N,M,j}$ 
and \perpx$^+_{N,M,j}$ with $j\geq 4$ are given, respectively, by
\begin{align}
 [j+4] - USp(N+Kj) &- SU(N+(K-1)j) -  \ldots - SU(N+2j) - SU(N+j) - [N]\,,
 \nonumber\\[1mm]
 [j-4] - \; SO(N+Kj)\; &- SU(N+(K-1)j) - \ldots - SU(N+2j) - SU(N+j) - [N]\,.
 \label{eq:perp-USp-SO-quivers}
\end{align}
The Chern-Simons levels are zero for all $SU(\cdot)$ nodes.
For the \perpx$^+_{N,M,j}$ theories $N+Kj$ can be odd.
The flavor bounds of \cite{Intriligator:1997pq} are saturated for all nodes.
The global symmetry manifest in the gauge theory deformation is $SO(2j+8)\times U(1)_I^K\times U(1)_B^{K}\times SU(N)$ for \perpx$^-_{N,2K,j}$
and $USp(2j-8)\times U(1)_I^K\times U(1)_B^{K}\times SU(N)$ for \perpx$^+_{N,2K,j}$.

For $M=2K+1$ odd there is a fractional NS5-brane on the O7, and the quivers for \perpx$^-_{N,M,j}$ 
and \perpx$^+_{N,M,j}$ are, respectively,
\begin{align}
[j+4]  - SU(N&+Kj) - SU(N+(K-1)j) - \ldots -  SU(N+2j) - SU(N+j) - [N]
\nonumber\\
 & \ \vert\nonumber\\
 &[1]^{}_A\nonumber\\[2mm]
[j-4] - SU(N&+Kj) - SU(N+(K-1)j) - \ldots -  SU(N+2j) - SU(N+j) - [N]
\nonumber\\
 & \ \vert\nonumber\\
 &[1]^{}_S
\label{eq:perp-S-A-quivers}
\end{align}
The Chern-Simons levels are zero for all nodes.
The global symmetry manifest in the gauge theory deformations for \perpx$_{N,2K+1,j}^-$ is $SU(j+4)\times U(1)_A\times U(1)_I^K\times U(1)_B^{K+1}\times SU(N)$.
For  \perpx$_{N,2K+1,j}^+$ with $j>4$ it is $SU(j-4)\times U(1)_S\times U(1)_I^K\times U(1)_B^{K+1}\times SU(N)$,
which reduces to $U(1)_S\times U(1)_I^K\times U(1)_B^{K}\times SU(N)$ for $j=4$.
As for the \vertx$_{N,j}^\pm$ theories, the nodes with (anti)symmetric matter are not in the classification of \cite{Intriligator:1997pq} if the rank is large.

Generically, the theories in (\ref{eq:perp-USp-SO-quivers}) and (\ref{eq:perp-S-A-quivers}) have a long meson operator, of the form discussed for other theories in \cite{Bergman:2018hin}.
Labeling the scalars in the bifundamental hypermultiplets represented by the links between adjacent gauge groups by $\lbrace x_i,\tilde x_i \rbrace$ with $i=1,\ldots,K-1$,
and labeling the scalars in the fundamental hypermultiplets at the first and last gauge node by $\lbrace y_0,\tilde y_0\rbrace$ and $\lbrace y_{K},\tilde y_{K}\rbrace$, respectively, the meson operator takes the form
\begin{align}\label{eq:perp-meson}
 \mathcal O&= y_0\cdot x_1\cdot \ldots \cdot x_{K-1}\cdot y_{K}~,
 &
 \Delta&=\frac{3}{2}(K+1)~.
\end{align}
This is a BPS operator with $R$-charge given by one third of its engineering dimension $\Delta$.
It exists when $N>0$ and there are fundamental hypermultiplets at the first gauge node, which is not the case for \perpx$_{N,M,j}^+$ with $j=4$.
In the string theory picture the meson operator is realized as a fundamental string connecting the external D5-branes to D7-branes coincident with the O7$^\pm$. This operator will be recovered in the supergravity solution as a consistency check.

\subsection{\texorpdfstring{\perpx$_{N,M,j}^\pm$}{perp[N,M,j]} solutions}

The supergravity solutions associated with the \perpx$^\pm_{N,M,j}$ junctions have two 5-brane poles and one puncture.
The solutions are identical for \perpx$^\pm_{N,M,j}$.
On the unit disc with coordinate $u$ they take the schematic form shown in fig.~\ref{fig:fund-perp}.
The functions $\cA_\pm$ are given by (\ref{eq:cApm-O7-only-u}) with $L_0=2$ and $I_0=1$.
The two regularity conditions in (\ref{eq:reg-O7-only}) are equivalent to each other and both reduce to
\begin{align}\label{eq:2-pole-reg}
 0&=Z^{[2,1]}\ln\left|\frac{t_1-t_2}{t_1+t_2}\right|^2+Y^1Y^2\frac{n_1^2}{2\pi}
 \left[
 \Li_2\left(\frac{t_1}{t_2}\right) -\Li_2\left(-\frac{t_1}{t_2}\right)-\rm{c.c.}
 \right].
\end{align}
The 5-brane and 7-brane charges of the \perpx$^\pm_{N,M,j}$ junctions are realized for
\begin{align}
 \cY_\pm^1&=\frac{3}{4}\alpha^\prime iN~,
 &
 \cY_\pm^2&=\pm\frac{3}{4}\alpha^\prime M~, & n_1^2=j~.
\end{align}
This implies that $Y^1=0$, and the second term drops out of the regularity condition (\ref{eq:2-pole-reg}).
Consequently, the regularity condition is solved for $t_2=it_1$.
As discussed above (\ref{eq:cApm-vert}), the position of one of the poles is a spurious parameter and one can fix
\begin{align}
 t_1&=1~, & t_2&=i~.
\end{align}
The branch cut associated with the puncture at $u=0$ intersects the pole at $u=1$, in line with the brane picture; as discussed in \cite{Gutperle:2018vdd} this does not affect the regularity conditions for D5-brane poles.
The residues at the poles are again given by $Z_\pm^\ell$, i.e.\ $\cY_\pm^\ell=Z_\pm^\ell$.
The resulting $\cA_\pm$ are
\begin{align}
 \cA_\pm&=\sum_{\ell=1}^2 \cY_\pm^\ell \ln\left(\frac{t_\ell-i\sqrt{-u}}{t_\ell+i\sqrt{-u}}\right)
 +\frac{j}{\pi}\cY_+^2
 \left(\Li_2(\sqrt{-u})-\Li_2(-\sqrt{-u})-\tanh^{-1}(\sqrt{-u})\ln(-u)\right).
\end{align}
The supergravity fields are obtained via (\ref{eqn:ansatz}) with (\ref{eq:kappa2-G}).

The meson operator (\ref{eq:perp-meson}) is realized by a fundamental string connecting the D5-brane pole to the O7/D7 puncture.
The fields that the string couples to are single-valued across the branch cut, and embedding it along the branch cut in fig.~\ref{fig:fund-perp} solves the equation of motion.
The scaling dimension of the operators represented by the string is extracted from the on-shell Hamiltonian (as discussed in detail in \cite{Bergman:2018hin}), and found as
\begin{align}
 \Delta_{\rm F1}&=\frac{3}{2}K~.
\end{align}
This agrees with the field theory result (\ref{eq:perp-meson}) for large $K$, providing a first check of the proposed duality.

With the supergravity dual the free energy of the \perpx$_{N,M,j}^\pm$ theories on $S^5$ can be computed using (\ref{eq:FS5-sugra-gen}).
The dependence on $N$, $M$ and $j$ can be found as follows.
The $\cA_\pm$ are linear in $j$. In $\partial_u\cG$ the quadratic terms in $j$ cancel, so that $\partial_u\cG$ is also linear in $j$ and  the free energy is a polynomial of degree 2 in $j$.
The quadratic term in $j$ scales like $M^4$, the term linear in $j$ scales like $M N^3$ and the $j=0$ term scales like $N^2M^2$.
This fixes the form of the free energy as function of $N$, $M$ and $j$.
The coefficients of the individual terms are determined by numerical integration, which yields
\begin{align}\label{eq:Oplus-FS5-sugra}
 F^{\text{\perpx$_{N,M,j}^\pm$}}_{S^5}&=
 -\frac{189}{32\pi^2}\zeta(3)N^2M^2 -0.430582\,j N M^3 -\frac{837}{128\pi^4}\zeta(5) j^2 M^4~.
\end{align}
For $j=0$ this is half the free energy of the $+_{N,M}$ theory (see e.g.\ \cite{Fluder:2018chf}).
For $N=0$ the result reduces to (\ref{eq:F-vert-sugra}) with $N\rightarrow M$.
For generic $N$ the free energy of the \perpx$_{N,M,j}^\pm$ theories is larger in absolute value than that of the \vertx$_{M,j}^\pm$ theories; since the \perpx$_{N,M,j}^\pm$ and \vertx$_{M,j}^\pm$ theories can be connected by an RG flow this is required for consistency with a possible 5d F-theorem.

\section{Free energies in field theory}\label{sec:F-loc}

In this section the free energies are discussed from the field theory perspective.
It will be shown that 5d SCFTs engineered by 5-brane webs with different O7/D7 combinations that lead to a puncture with the same $-T^{k_0}$ monodromy in supergravity have identical free energies, which are related to the free energy of 5d SCFTs that are engineered by 5-brane webs without O7-plane.
The free energies of the \vertx$_{N,j}^\pm$ and \perpx$_{N,M,j}^\pm$ theories of sec.~\ref{sec:5dSCFTs} will be obtained explicitly.

The general forms of the quiver theories obtained from 5-brane junctions on a combination of an O7$^\pm$ with D7-branes leading to $-T^{k_0}$ monodromy 
(and possibly further D7-branes) are
\begin{align}\label{eq:SU-A-quiver}
[1]^{}_A- SU(&N_0) - SU(N_1)_{c^{}_1} - \ldots  - SU(N_L)_{c^{}_L}
 \nonumber\\[-1mm]
 &\vert \hskip 16mm \vert \hskip 30mm \vert
  \\[-1mm] \nonumber
  & \!\!\!\!\!\!\!\!\!\! [k_0+4] \hskip 9mm [k_1] \hskip 25mm [k_L]
\\[3mm]
  [1]^{}_S- SU(&N_0) - SU(N_1)_{c^{}_1} - \ldots  - SU(N_L)_{c^{}_L}
 \nonumber\\[-1mm]
 &\vert \hskip 16mm \vert \hskip 30mm \vert
 \label{eq:SU-S-quiver}
  \\[-1mm] \nonumber
  & \!\!\!\!\!\!\!\!\!\! [k_0-4] \hskip 9mm [k_1] \hskip 25mm [k_L]
\end{align}
for 5-brane webs with a fractional NS5-brane on the O7-plane, and 
\begin{align}
\label{eq:USp-quiver}
 USp(&N_0) - SU(N_1)_{c^{}_1} - \ldots  - SU(N_L)_{c^{}_L}
 \nonumber\\[-1mm]
 &\vert \hskip 16mm \vert \hskip 30mm \vert
  \\[-1mm] \nonumber
  & \!\!\!\!\!\!\!\!\!\! [k_0+4] \hskip 9mm [k_1] \hskip 25mm [k_L]
\\[3mm]
\label{eq:SO-quiver}
 SO(&N_0) - SU(N_1)_{c^{}_1} - \ldots  - SU(N_L)_{c^{}_L}
 \nonumber\\[-1mm]
 &\vert \hskip 16mm \vert \hskip 30mm \vert
  \\[-1mm] \nonumber
  & \!\!\!\!\!\!\!\!\!\! [k_0-4] \hskip 9mm [k_1] \hskip 25mm [k_L]
\end{align}
for 5-brane webs with no fractional NS5-brane.
The subscripts $c_t$, if present, denote Chern-Simons levels.
5-brane webs on an O7$^-$ with $k_0+4$ coincident D7-branes realize quivers of the form (\ref{eq:SU-A-quiver}), (\ref{eq:USp-quiver}). 5-brane webs on an O7$^+$ with $k_0-4\geq 0$ D7-branes realize quivers of the form (\ref{eq:SU-S-quiver}), (\ref{eq:SO-quiver}).

It will be shown in field theory that the free energies of the UV SCFTs of the general quivers above, if they are of the type with $L\gg 1$ which has holographic duals in Type IIB, are identical, and given by half the free energy of
\begin{align}\label{eq:unfolded-quiver}
 SU(&N_L)_{-c^{}_L} - \ldots - SU(N_1)_{-c^{}_1} - SU(N_0)_0 - SU(N_1)_{c^{}_1} - \ldots  - SU(N_L)_{c^{}_L}
 \nonumber \\[-1mm]
 &\,\vert \hskip 33mm \vert \hskip 23mm \vert \hskip 18mm \vert \hskip 30mm \vert
 \\[-1mm] \nonumber
 & \!\!\! [k_L] \hskip 28mm [k_1] \hskip 17mm [2k_0] \hskip 13mm [k_1] \hskip 25mm [k_L]
\end{align}
This is the field theory version of the statement that the supergravity solutions corresponding to the different O7/D7/NS5 combinations are identical, thus leading to the same free energies, and that they are obtained as $\ZZ_2$ quotient of solutions without O7 puncture.\footnote{%
This is reminiscent of the situation for orbifolds of the $USp(N)$ theories \cite{Bergman:2012kr,Jafferis:2012iv}.
The gauge theories here are quivers to begin with, with a large number of parameters in the parent theories and a different scaling of the free energies. But the universal relation between the free energies of the parent and descendant theories is analogous.}

The argument will proceed as follows. 
The free energies of the 5d SCFTs are computed using supersymmetric localization in the gauge theory deformations.
The ``bulk'' of the quivers (\ref{eq:SU-A-quiver}) -- (\ref{eq:SO-quiver}), i.e.\ the set of interior gauge nodes, consists of $SU(\cdot)$ nodes with only fundamental matter and possibly Chern-Simons terms. 
The matrix models computing the partition functions for such quivers and their saddle point evaluation capturing the free energy at large $L$ were discussed in \cite{Uhlemann:2019ypp}.
The (bulk) requirement for non-trivial saddle points fixes the scaling of the eigenvalues to be linear in $L$,
and the saddle point equations take the form of an electrostatics problem with certain boundary conditions.
Since the bulk for the quivers (\ref{eq:SU-A-quiver}) -- (\ref{eq:SO-quiver}) is 
of the form discussed in \cite{Uhlemann:2019ypp}, the arguments for the scaling of the eigenvalues apply.
The modified boundary nodes can thus be incorporated by adding and manipulating the leading-order contributions of the boundary nodes to the matrix models.
For the quivers (\ref{eq:USp-quiver}), (\ref{eq:SO-quiver}) this directly leads to the identification with (\ref{eq:unfolded-quiver}),
for (\ref{eq:SU-A-quiver}), (\ref{eq:SU-S-quiver}) it follows upon closer inspection of the modified boundary conditions for the saddle point equations.
Since the quiver (\ref{eq:unfolded-quiver}) is of the type discussed in \cite{Uhlemann:2019ypp}, this identification allows to derive explicit results for all five related theories.

The general form of the matrix models resulting from supersymmetric localization on squashed $S^5$ and the formulation for long quiver theories are reviewed in sec.~\ref{sec:loc-rev}.
The quivers (\ref{eq:SU-A-quiver}), (\ref{eq:SU-S-quiver}) and (\ref{eq:USp-quiver}), (\ref{eq:SO-quiver}) are discussed in sec.~\ref{sec:SU-A-S-loc} and sec.~\ref{sec:USp-SO-loc}, respectively.
Results for the \vertx$_{N,j}^\pm$ and \perpx$^\pm_{N,M,j}$ theories are given in sec.~\ref{sec:free-energy-1}.

\subsection{Long quivers}\label{sec:loc-rev}

The perturbative part of the squashed $S^5$ partition function for a 5d gauge theory with gauge group $G$ and $N_f$ hypermultiplets in a real representation $R_f \otimes \bar R_{f}$ of $G$ is given by \cite{Lockhart:2012vp,Imamura:2012xg,Imamura:2012bm}
\begin{align}\label{eqn:generalpartfunc}
\cZ_{\vec{\omega}}  &= 
\mathcal C(\vec{\omega})
\left[ \prod_{i=1}^{\mathrm{rk} \, G} \int_{-\infty}^{\infty} d \lambda_i \right]e^{-\frac{(2\pi)^{3}}{\omega_1\omega_2\omega_3}\mathfrak{F}(\lambda)}
\times 
\frac{\prod_{\alpha} S_3 \left( -i \alpha(\lambda) \mid \vec{\omega} \right)}{\prod_{f=1}^{N_f} \prod_{\rho_{f}} S_{3} \left( i \rho_{f}(\lambda) + \tfrac{\omega_{\rm tot}}{2} \mid \vec{\omega}\right)} ~.
\end{align}
The squashing parameters are encoded in $\vec{\omega}=(\omega_1,\omega_2,\omega_3)$ with $\omega_{\rm tot}\equiv\omega_1+\omega_2+\omega_3$, the round $S^5$ corresponds to $\vec{\omega}=(1,1,1)$.
$S_3 (z \mid \vec{\omega})$ is the triple sine function.
The roots of $G$ are denoted by $\alpha$; $f = 1, \ldots N_f$ labels the flavor hypermultiplets and $\rho_f$ are the weights of the corresponding representation. 
$\mathfrak{F}(\lambda)$ is the classical flat space prepotential, which for $g_{\rm YM}\rightarrow\infty$ reduces to Chern-Simons terms.
Finally, $\mathcal C(\vec{\omega})=\left| \cW \right|^{-1}S_{3}^{\prime} \left( 0\mid \vec{\omega} \right)^{\mathrm{rk} \, G}/(2\pi)^{\mathrm{rk} \, G}$
with $\cW $ the Weyl group of $G$.
The contributions due to vector multiplets and hypermultiplets are collected, respectively, in
\begin{align}
 F_V(x)&\equiv-\omega_1\omega_2\omega_3\frac{\ln S_3\left(ix|\vec{\omega}\right)+\ln S_3\left(-ix|\vec{\omega}\right)}{2}
 \approx 
 \frac{\pi}{6}|x|^3
 -\frac{\omega_{\rm tot}^2+\omega_1 \omega_2+\omega_1\omega_3+\omega_2\omega_3}{12}\pi|x|~,
\nonumber\\
F_H(x)&\equiv \omega_1\omega_2\omega_3\ln S_3\left(i x+\frac{\omega_{\rm tot}}{2}\mid \vec{\omega}\right)
\approx-\frac{\pi}{6}|x|^3-\frac{\omega_1^2+\omega_2^2+\omega_3^2}{24}\pi|x|~.
\end{align}
Only the behavior for large argument will be relevant in the following.

The partition functions and free energies $F_{\vec{\omega}}=-\ln\cZ_{\vec{\omega}}$ for quiver theories of the general form
\begin{align}\label{eq:SU-fund-quiver}
SU(&N_0)_{c^{}_0} - SU(N_1)_{c^{}_1} - \ldots  - SU(N_L)_{c^{}_L}
 \nonumber\\[-1mm]
 &\vert \hskip 20mm \vert \hskip 29mm \vert
  \\[-1mm] \nonumber
  & \!\!\!\! [k_0] \hskip 15mm [k_1] \hskip 24.5mm [k_L]
\end{align}
with $L\gg 1$ were discussed in \cite{Uhlemann:2019ypp}.
With $\lbrace e_i\vert i=1,..,N\rbrace$ the weights of the fundamental representation of $SU(N)$
and $\lbrace e_i-e_j\vert  i\neq j\rbrace$ the roots,
the partition function is
\begin{align}\label{eq:cF-fund}
 \cZ_{\vec{\omega}} &=  \mathcal C(\vec{\omega})
\left[ \prod_{i=1}^{\mathrm{rk} \, G} \int_{-\infty}^{\infty} d \lambda_i \right] e^{-\cF_{\vec{\omega}}}\,,
&
 \cF_{\vec{\omega}} &= 
 \sum_{t=0}^L\sum_{\stackrel{\ell,m=1}{\ell\neq m}}^{N_t}F_V\big(\lambda_\ell^{(t)}-\lambda_m^{(t)}\big)
 +\sum_{t=0}^{L-1}\sum_{\ell=1}^{N_t}\sum_{m=1}^{N_{t+1}}F_H\big(\lambda_\ell^{(t)}-\lambda_m^{(t+1)}\big)
 \nonumber\\ &&&
 \hphantom{=} +\sum_{t=0}^L\sum_{\ell=1}^{N_t} \left(k_tF_H\big(\lambda_\ell^{(t)}\big)+\frac{\pi}{3}c_t\left(\lambda_\ell^{(t)}\right)^3\right)
 ~.
\end{align}
A normalized eigenvalue distribution for each node is introduced as $\rho_t(\lambda)$, $t=1,\ldots,L$.
For large $L$ the quiver (\ref{eq:SU-fund-quiver}) is more appropriately described by continuous data
\begin{align}\label{eq:cont-quiver}
  z&=\frac{t}{L}~, &   N(z)&= N_{z L}~,  & k(z)&= k_{zL}~, & c(z)&=c_{zL}~,
\end{align}
and the eigenvalue distributions are captured by one function of two variables, $\rho(z,\lambda)=\rho_{zL}(\lambda)$.
For the theories of interest $N(z)$ is a piecewise linear and concave function with values of order $L$, 
and fundamental hypermultiplets and Chern-Simons terms appear only at the kinks of $N(z)$.
The free energy at large $L$ is captured by a saddle point evaluation, and the requirement for non-trivial 
saddle points to exist fixes the scaling of the eigenvalues to be linear in $L$.
The eigenvalues can thus be parametrized by an order-one parameter $x$, 
and a rescaled distribution function $\varrho$ is introduced as follows,
\begin{align}
 \lambda&\equiv \omega_{\rm tot} L x~, &
 \varrho(z,x)&=N(z)L\omega_{\rm tot}\rho(z,L\omega_{\rm tot}x)~.
\end{align}
The normalization of $\varrho(z,x)$ is $N(z)$, which is typically of order $L$.
Discrete sums over eigenvalues and quiver nodes are replaced as follows,
\begin{align}\label{eq:rho}
 \sum_{\ell=1}^{N_t} f(\lambda_\ell^{(t)})& 
 \ \longrightarrow \
 \int dx\, \varrho(z,x) f(\lambda)~,
 &
 \sum_{t=1}^{L} f_t \ &\longrightarrow \ L\int_0^1 dz f(z L)~.
\end{align}
With the continuous variables and parameters the large-$L$ partition function becomes
\begin{align}
 \cZ_{\vec{\omega}} \ = \ &
 \int \mathcal D \varrho\, \exp\left(-\frac{\omega_{\rm tot}^3}{\omega_1\omega_2\omega_3}\cF\right)~,
 \nonumber\\
  \cF \ = \ &
 \int_0^1 dz \left[L^2\int dx\,dy\,\cL+L^4\int dx\, \varrho(z,x)\left(k(z)F_H(x)+\frac{\pi}{3}c(z)x^3\right)\right]
 \nonumber\\
 &
 -\frac{1}{2}L^3\sum_{z\in\lbrace 0,1\rbrace}\int dx\,dy\,\varrho(z,x)\varrho(z,y)F_H\big(x-y\big)
 \nonumber\\ 
 \cL \ = \ &
 \varrho(z,x)\varrho(z,y)F_0(x-y\big)
 -\frac{1}{2}\partial_z \varrho(z,x)\partial_z\varrho(z,y)F_H\big(x-y\big)~,
 \label{eq:cZ-cont}
\end{align}
where $F_0(x)\equiv (F_V(x)+F_H(x))/\omega_{\rm tot}^2$.
The saddle point equations take the form of an electrostatics problem, with sources provided by fundamental hypermultiplets and  boundary and junction conditions depending on the shape of the quiver and Chern-Simons terms.
More details can be found in \cite{Uhlemann:2019ypp} (Lagrange multiplier terms are suppressed here to keep the notation simple).

\subsection{Quivers with (anti)symmetric matter}\label{sec:SU-A-S-loc}

The difference in the quivers (\ref{eq:SU-A-quiver}), (\ref{eq:SU-S-quiver}) compared to (\ref{eq:SU-fund-quiver}) is the presence of (anti)symmetric hypermultiplets and different numbers of fundamental hypermultiplets at the $t=0$ node.
With the weights of the antisymmetric representation $\lbrace e_i+e_j\vert i<j\rbrace$ and of the symmetric $\lbrace e_i+e_j\vert i\leq j\rbrace$, the exponent in (\ref{eq:cF-fund}) for the quivers (\ref{eq:SU-A-quiver}) and (\ref{eq:SU-S-quiver}) is, respectively,
\begin{align}
 \cF_{\vec{\omega}}^A&=\cF_{\vec{\omega}}+4\sum_{\ell=1}^{N_0} F_H\big(\lambda_\ell^{(0)}\big)
 +\sum_{\ell<m}F_H\big(\lambda^{(0)}_\ell+\lambda^{(0)}_m\big)~,
 \nonumber\\
 \cF_{\vec{\omega}}^S&=\cF_{\vec{\omega}}-4 \sum_{\ell=1}^{N_0} F_H\big(\lambda_\ell^{(0)}\big)
 +\sum_{\ell\leq m}F_H\big(\lambda^{(0)}_\ell+\lambda^{(0)}_m\big)~,
\end{align}
with $\cF_{\vec{\omega}}$ as given in (\ref{eq:cF-fund}).
Using that the leading term in $F_H$ is cubic in the argument and depends only on the absolute value, both cases reduce to
\begin{align}\label{eq:A-S-bndy-cF-3}
 \cF_{\vec{\omega}}^{A/S}&=\cF_{\vec{\omega}}+\Delta\cF_{\vec{\omega}}~,
 &
 \Delta\cF_{\vec{\omega}}&= \frac{1}{2}\sum_{\ell=1}^{N_0}\sum_{m=1}^{N_0}F_H\big(\lambda^{(0)}_\ell+\lambda^{(0)}_m\big)~.
\end{align}
For the symmetric hypermultiplet the diagonal contribution is equivalent to $4$ fundamentals,
for the antisymmetric the contribution of $4$ fundamental hypermultiplets has been absorbed.
In the continuous version (\ref{eq:cZ-cont}), using the replacements (\ref{eq:cont-quiver}), (\ref{eq:rho}) and the rescaling of $\cF_{\vec{\omega}}$ to absorb the dependence on the squashing parameters, this results in
\begin{align}
 \cF&\rightarrow \cF+\Delta \cF~,
 &
 \Delta \cF&=
 \frac{1}{2}L^3 \int dx\,dy\,\varrho(0,x)\varrho(0,y)F_H(x+y)~.
\end{align}

The additional term affects the boundary conditions at $z=0$.
They are derived from a variation of $\varrho(z,x)$ with support in the neighborhood of $z=0$, assuming that the (unmodified) bulk equation of motion is satisfied away from $z=0$.
This yields, using $k(z)=L^{-1}k_0\delta(z)+\ldots$
\begin{align}\label{eq:delta-cF-AS}
 \delta(\cF+\Delta \cF) = 
 \int dx\,\delta \varrho(0,x)\Bigg[&
 L^3\int dy\,\varrho(0,y)\left(F_H(x+y)-F_H(x-y)\right)
 \nonumber\\ &
 +L^3k_0 F_H(x)
 +L^2\int dy\,\partial_z\varrho(z,y)\Big\vert_{z=0}F_H(x-y)\Bigg]
 ~.
\end{align}
The resulting boundary condition reads
\begin{align}\label{eq:bc-AS-2}
 0 &= \int dy\Big[L\varrho(0,y)\left(F_H(x+y)-F_H(x-y)\right)
 +\partial_z\varrho(z,y)\big\vert_{z=0}F_H(x-y)+Lk_0\delta(y)F_H(x)\Big]\,.
\end{align}
At the node with the (anti)symmetric, the rank of the gauge group, $N(0)$, is of order $L$ (that is, this node does not correspond to a quiver tail along which the gauge group decreases to order one; this is clear from the brane construction).
Noting that $\varrho(z,x)$ is of order $N(z)$, the first term in the square brackets is the dominant contribution at large $L$.
Acting with $\partial_x^4$ on (\ref{eq:bc-AS-2}) and keeping only the leading contribution leads to\footnote{%
Even though the $t=0$ node in (\ref{eq:SU-A-quiver}), (\ref{eq:SU-S-quiver}) does not have a Chern-Simons term, $\cF$ is in general not invariant under $\varrho(0,x)\rightarrow \varrho(0,-x)$. Due to the coupling between eigenvalue distributions at adjacent nodes in the derivative terms in $\cL$, any node with a Chern-Simons term breaks the symmetry under reflection of the eigenvalues for all nodes, as exhibited in the examples of \cite{Uhlemann:2019ypp}.}
\begin{align}\label{eq:bc-AS-1}
 \varrho(0,x)&=\varrho(0,-x)~.
\end{align}
With this constraint the $\mathcal O(L^5)$ boundary terms in (\ref{eq:delta-cF-AS}) and the first term in (\ref{eq:bc-AS-2}) vanish.
The actual boundary condition for $\varrho$ is then given by the vanishing of the remaining terms in (\ref{eq:bc-AS-2}),
which leads to a Neumann type boundary condition and has to be combined with (\ref{eq:bc-AS-1}).
For the quiver (\ref{eq:SU-fund-quiver}) the boundary condition is Dirichlet, as discussed in detail in \cite{Uhlemann:2019ypp}.
The effect of the extra (anti)symmetric hypermultiplet therefore is to turn the Dirichlet type boundary condition into a Neumann type boundary condition.

It is now shown that the quiver (\ref{eq:unfolded-quiver}) leads to twice the free energy,
which can be done without explicitly spelling out the saddle point equations.
The exponent $\widehat\cF_{\vec{\omega}}$ for the quiver (\ref{eq:unfolded-quiver}) is
\begin{align}\label{eq:cF-gen-unfolded}
\widehat\cF_{\vec{\omega}} \ = \ &
 \sum_{t=-L}^{+L}\sum_{\stackrel{\ell,m=1}{\ell\neq m}}^{N_t}F_V\big(\lambda_\ell^{(t)}-\lambda_m^{(t)}\big)
 +\sum_{t=-L}^{L-1}\sum_{\ell=1}^{N_t}\sum_{m=1}^{N_{t+1}}F_H\big(\lambda_\ell^{(t)}-\lambda_m^{(t+1)}\big)
 \nonumber\\ &
   +\sum_{\stackrel{t=-L}{t\neq 0}}^L\sum_{\ell=1}^{N_t} \left(k_tF_H\big(\lambda_\ell^{(t)}\big)+\frac{\pi}{3}c_t\left(\lambda_\ell^{(t)}\right)^3\right)
   + 2k_0\sum_{\ell=1}^{N_0}F_H\big(\lambda_\ell^{(0)}\big)
 ~.
\end{align}
Since the quiver (\ref{eq:unfolded-quiver}) is symmetric under reflection across the central node combined with charge conjugation,
one expects a saddle point with the same symmetry. 
That is, the saddle point is expected to be invariant under
 \begin{align}\label{eq:unfolded-sym-1}
  \lambda_\ell^{(-t)}&\rightarrow -\lambda_{\ell}^{(t)}~, \qquad t=0,\ldots, L~.
 \end{align}
Using (\ref{eq:unfolded-sym-1}) to fold the sums in (\ref{eq:cF-gen-unfolded}), to restrict to $t\geq 0$, 
leads to
\begin{align}\label{eq:cF-unfolded-4}
 \widehat\cF_{\vec{\omega}}&=2\cF_{\vec{\omega}}
 -\sum_{\stackrel{\ell,m=1}{\ell\neq m}}^{N_0}F_V\big(\lambda_\ell^{(0)}-\lambda_m^{(0)}\big)
 ~,
\end{align}
with $\cF_{\vec{\omega}}$ as given in (\ref{eq:cF-fund}).
The integration variables in the folded expression are constrained by (\ref{eq:unfolded-sym-1}) with $t=0$.
Comparing to (\ref{eq:A-S-bndy-cF-3}), using the constraint in (\ref{eq:unfolded-sym-1}) and that at leading order for large arguments $F_H(x)$ and $F_V$ are opposite-equal, one finds 
\begin{align}
\widehat\cF_{\vec{\omega}}&=2\cF_{\vec{\omega}}^{A/S}~.
\end{align}
The constraint (\ref{eq:bc-AS-1}) for $\cF_{\vec{\omega}}^{A/S}$ corresponds to (\ref{eq:unfolded-sym-1})  with $t=0$.
The free energies are therefore derived by extremizing the same functions, up to a factor $2$, subject to the same constraints,
showing  that they are related by a factor $2$.

From the point of view of the electrostatics problem from which the saddle point distributions $\varrho(z,x)$ are determined \cite{Uhlemann:2019ypp}, the Neumann boundary condition resulting from (\ref{eq:bc-AS-2}), (\ref{eq:bc-AS-1}) can be implemented by introducing mirror charges, obtained by reflection through the point $x=z=0$.
In (\ref{eq:cZ-cont}) this corresponds to extending $N(z)$, $k(z)$, $c(z)$ and $\rho(z,x)$ to $z\in [-1,1]$, with $N(z)$ and $k(z)$ even and $c(z)$ odd. To realize the junction condition (\ref{eq:bc-AS-1}) the number of flavors at $z=0$ has to be doubled.
In gauge theory terms this corresponds to unfolding the quivers (\ref{eq:SU-A-quiver}), (\ref{eq:SU-S-quiver}) with the Chern-Simons levels reversed in the mirror part, leading to (\ref{eq:unfolded-quiver}).

\subsection{Quivers with \texorpdfstring{$USp(\cdot)$}{USp()} and \texorpdfstring{$SO(\cdot)$}{SO()} nodes}\label{sec:USp-SO-loc}

The quivers (\ref{eq:USp-quiver}), (\ref{eq:SO-quiver}) differ from (\ref{eq:SU-fund-quiver}) in the type of the first node.
It will be convenient to separate the contribution from the first node and bifundamentals between the first and second node from the remaining part, and write the exponent in (\ref{eq:cF-fund}) as
\begin{align}
\cF_{\vec{\omega}}^{USp/SO}&=\cF_{t=0}^{USp/SO}+\cF_{t>0}~.
\end{align}
$\cF_{t>0}$ takes the form given by $\cF_{\vec{\omega}}$ in (\ref{eq:cF-fund}) except that all sums over $t$ start at $t=1$. 
The contribution $\cF_{t=0}$ will be discussed now.
With $\lbrace \pm e_i\vert i=1,..,N\rbrace$ the weights of the fundamental representation of $USp(2N)$
and $\lbrace \pm e_i\pm e_j \vert i,j=1,\ldots,N, \ i\leq j\rbrace$ the roots,
for the quiver (\ref{eq:USp-quiver}) with $N_0$ even
\begin{align}\label{eq:cF-gen-USp}
 \cF_{t=0}^{USp}  &= 
 \sum_{\stackrel{\ell,m=1}{\ell\leq m}}^{N_0/2}F_V\big(\pm \lambda_\ell^{(0)}\pm\lambda_m^{(0)}\big)
 +(k_0+4)\sum_{\ell=1}^{N_0/2}F_H(\pm\lambda_\ell^{(0)})
 +\sum_{\ell=1}^{N_0/2}\sum_{m=1}^{N_{1}}F_H\big(\pm\lambda_\ell^{(0)}-\lambda_m^{(1)}\big)
 ~.
\end{align}
The sign choices in the arguments of $F_V$ and $F_H$ are to be summed over separately.
The roots of $SO(2N)$ and $SO(2N+1)$ are $\lbrace \pm e_i\pm e_j \vert i,j=1,..,N, \ i<j\rbrace$, and the weights of the vector representation are $\lbrace \pm e_i\vert i=1,..,N\rbrace$.
For the quiver in (\ref{eq:SO-quiver}),
\begin{align}\label{eq:cF-gen-SO}
  \cF_{t=0}^{SO}  &= 
 \sum_{\stackrel{\ell,m=1}{\ell< m}}^{N_0/2}F_V\big(\pm \lambda_\ell^{(0)}\pm\lambda_m^{(0)}\big)
 +(k_0-4)\sum_{\ell=1}^{N_0/2}F_H(\pm\lambda_\ell^{(0)})
 +\sum_{\ell=1}^{N_0/2}\sum_{m=1}^{N_{1}}F_H\big(\pm\lambda_\ell^{(0)}-\lambda_m^{(1)}\big)
 ~.
\end{align}
For both cases it is convenient to double the number of integration variables at the expense of introducing a constraint.
Namely, define $\lambda_{N_0/2+1},\ldots,\lambda_{N_0}$ by
\begin{align}\label{eq:USp-constr}
 \lambda_{i+N_0/2}^{(0)}&=-\lambda_i^{(0)}~, \qquad i=1,\ldots N_0/2~.
\end{align}
The sums over the sign choices in (\ref{eq:cF-gen-USp}), (\ref{eq:cF-gen-SO}) are then implemented by summing over the extended integration variables.
Moreover, since $F_H$ and $F_V$ are both cubic at leading order with opposite-equal coefficients, 
the $\ell=m$ terms in the first sums can be combined with the second terms.
As a result, in both cases,
\begin{align}\label{eq:cF-gen-USp-2}
\cF_{t=0}^{USp/SO}&= 
 \frac{1}{2}\sum_{\ell,m=1}^{N_0}F_V\big(\lambda_\ell^{(0)}-\lambda_m^{(0)}\big)
 +k_0\sum_{\ell=1}^{N_0}F_H(\lambda_\ell^{(0)})
 +\sum_{\ell=1}^{N_0}\sum_{m=1}^{N_{1}}F_H\big(\lambda_\ell^{(0)}-\lambda_m^{(1)}\big)
 ~.
\end{align}
With the constraint (\ref{eq:USp-constr}) this is equivalent to
\begin{align}\label{eq:cF-gen-USp-3}
\cF_{\partial}^{USp/SO} &= 
 \sum_{\ell,m=1}^{N_0}F_V\big(\lambda_\ell^{(0)}-\lambda_m^{(0)}\big)
 +k_0\sum_{\ell=1}^{N_0}F_H(\lambda_\ell^{(0)})
 +\sum_{\ell=1}^{N_0}\sum_{m=1}^{N_{1}}F_H\big(\lambda_\ell^{(0)}-\lambda_m^{(1)}\big)
 + \Delta\cF_{\vec{\omega}}
 \,,
\end{align}
with $\Delta\cF_{\vec{\omega}}$ as defined in (\ref{eq:A-S-bndy-cF-3}).
Comparing to the $t=0$ terms in (\ref{eq:cF-fund}) shows that
\begin{align}
 \cF_{\vec{\omega}}^{USp/SO}&=\cF_{\vec{\omega}} + \Delta\cF_{\vec{\omega}}~.
\end{align}
This is half the exponent for the quiver (\ref{eq:unfolded-quiver}), as given in (\ref{eq:cF-unfolded-4}).
The constraint (\ref{eq:USp-constr}) corresponds to (\ref{eq:unfolded-sym-1}) with $t=0$.
The free energies are thus obtained by extremizing the same functions, up to a factor $2$, with the same constraints,
showing that they are related by a factor $2$.

\subsection{\texorpdfstring{\vertx$_{N,j}^\pm$}{I[N,j]} and \texorpdfstring{\perpx$^\pm_{N,M,j}$}{perp[N,M,j]} theories}\label{sec:free-energy-1}

The results of the previous sections show that the free energy on squashed $S^5$ for the orientifold quivers is generally related by a factor $2$ to that of parent theories of the form (\ref{eq:unfolded-quiver}), which is a special case of (\ref{eq:SU-fund-quiver}).
The free energies for these parent theories can be obtained using the methods of \cite{Uhlemann:2019ypp}.
Unfolding the quivers (\ref{eq:I-USp-SO-quivers}), (\ref{eq:I-SU-A-S-quivers}) for the \vertx$_{N,j}^\pm$ theories and (\ref{eq:perp-USp-SO-quivers}), (\ref{eq:perp-S-A-quivers}) for the \perpx$_{N,M,j}^\pm$ theories leads to quivers of the form (\ref{eq:SU-fund-quiver}) with $N_f=2N_c$ at each node. 
The free energy for such theories was already derived explicitly in \cite{Uhlemann:2019ypp}. 
For the round $S^5$ the general expression is
\begin{align}\label{eq:cF-gen-Nf2NC}
 F_{S^5} = \, &
 -\frac{27L^2}{16\pi^2}\left[2N_0^2+2N_L^2+3N_0N_L\right]\zeta(3)
 -\frac{27L^3}{4\pi^3}\sum_{t=1}^{L-1}k_t 
 \left[N_0 D_4\big(e^{i \pi  z_t}\big)+N_L D_4\big(e^{i \pi (1- z_t)}\big)\right]
 \nonumber\\ &
 +\frac{27L^4}{16 \pi^4}\sum_{t=1}^{L-1}\sum_{s=1}^{L-1}k_tk_s\left[D_5\big(e^{i\pi(z_s+z_t)}\big)-D_5\big(e^{i\pi(z_s-z_t)}\big)\right]~,
\end{align}
with $D_4(e^{i\alpha})=\Im({\rm Li}_4(e^{i\alpha}))$ and $D_5(e^{i\alpha})=\Re({\rm Li}_5(e^{i\alpha}))$.

The unfolded version of the $\vertx_{N,j}^\pm$ theories has $N_0=N_L=0$ and $2j$ flavors at $z=\frac{1}{2}$.
The length of the unfolded quiver is $L=N+\mathcal O(1)$. Thus,
\begin{align}\label{eq:Oplus-F}
 F^{\vertx_{N,j}^\pm}_{S^5}&=-\frac{837}{128\pi^4}\zeta (5)j^2N^4~,
\end{align}
in agreement with (\ref{eq:F-vert-sugra}).
For the unfolded version of the \perpx$^\pm_{N,M,j}$ theories, $N_0=N_L=N$ and $L=M$, with again $2j$ flavors at $z=\frac{1}{2}$.
Thus,
\begin{align}
  F_{S^5}^{\text{\perpx}^\pm_{N,M,j}}&=-\frac{27}{2\pi^2}M^2\left(\frac{7}{16}\zeta(3)N^2+\frac{D_4(i)}{\pi}jNM+\frac{31}{64\pi^2}\zeta(5)j^2M^2\right)~.
\end{align}
This agrees with (\ref{eq:Oplus-FS5-sugra}) and provides the analytic form for the numerical coefficients.
These results provide an explicit match between supergravity results and field theory computations, to support the proposed dualities between supergravity solutions with O7 planes and field theories engineered by 5-brane junctions on O7 planes.

\section{Discussion}\label{sec:discussion}

$AdS_6$ solutions to Type IIB supergravity were constructed which incorporate orientifold 7-planes in addition to $(p,q)$ 5-branes and D7-branes. They were obtained as quotients of certain limiting cases of solutions with D7-branes.
The geometry is a warped product of $AdS_6$ and $S^2$ over a Riemann surface, which contains a puncture with $-T^j$ monodromy.
The solutions were identified as near-horizon limits of 5-brane webs on either an O7$^-$ with $j+4$ D7-branes, or, for $j\geq 4$, on an O7$^+$ with $j-4$ D7-branes, and provide holographic duals for the 5d SCFTs engineered by such 5-brane webs. 
The general ansatz allows to construct solutions for large classes of 5-brane webs on O7-planes, 
and the parameters of the solutions were identified with those of the 5-brane webs.
Two simple classes of solutions were discussed in detail.
Regularity of the supergravity solutions requires $j\geq 0$, which includes isolated O7$^+$ for $j=4$ but excludes isolated O7$^-$.
Whether the regularity conditions can be relaxed without sacrificing the clear holographic interpretation of the solutions is an interesting question left for the future (a recent general discussion of issues relating to O$p^-$ planes in supergravity can be found in \cite{Cordova:2019cvf}).

The 5d SCFTs for which the solutions constructed here provide holographic duals have relevant deformations to quiver gauge theories with a large number of $SU(\cdot)$ nodes, which may involve Chern-Simons terms and fundamental hypermultiplets. 
Due to the localized O7-planes they also involve, depending on the type of O7 plane and the involved 5-branes, $USp(\cdot)$ or $SO(\cdot)$ gauge nodes, or $SU(\cdot)$ nodes with matter in the symmetric or antisymmetric representation.
Two simple classes of 5-brane junctions on O7-planes and the associated 5d SCFTs were discussed in detail, along with the supergravity solutions that provide holographic duals for the 5d SCFTs.
The holographic duals were used to compute the $S^5$ free energies, which only depend on the monodromy of the O7/D7 combination and not on its precise realization in string theory.
How the $-T^j$ monodromy is realized in the string theory construction does affect the spectrum of string and brane states on the solutions, and thus the spectrum of gauge invariant operators, as exemplified by the stringy meson operator in the \perpx$_{N,M,j}^\pm$ theory. It would be interesting to study the spectrum more comprehensively.

In the last part the free energies were studied in field theory using supersymmetric localization.
It was shown that, as far as the free energy of long quiver gauge theories whose UV fixed points have holographic duals in Type IIB is concerned, the following are equivalent:
a $USp(N)$ node with $8$ fundamental hypermultiplets, an $SO(N)$ node, an $SU(N)$ node with an antisymmetric and 8 fundamental hypermultiplets, and an $SU(N)$ node with a hypermultiplet in the symmetric representation.
This established in field theory that the free energies are identical for the different string theory realizations of the supergravity solutions, i.e.\ whether the $-T^j$ monodromy is realized by an O7$^-$ with $4+j$ D7-branes or by an O7$^+$ with $j-4$ D7-branes, and whether there is a fractional NS5 on the O7 or not.
It was also shown that the free energy of the orientifold theories is given by half the free energy of a parent theory with only $SU(\cdot)$ nodes and fundamental matter, in line with the supergravity construction.
For the SCFTs discussed in the second part explicit expressions for the free energies were obtained,
which exactly match the supergravity predictions.

Several interesting questions are left for the future. 
For example, for $AdS_6$ solutions without punctures on the Riemann surface a connection to the M-theory construction of 4d $\mathcal N=2$ SCFTs was established in \cite{Kaidi:2018zkx}, and it would be interesting to extend this relation to solutions with punctures.
Another question is for further generalization of the solutions. For example, 5-brane junctions with O5-planes allow for the construction of interesting SCFTs \cite{Zafrir:2015ftn,Hayashi:2018bkd,Hayashi:2018lyv,Hayashi:2019yxj}, and their study would likely benefit from the existence of holographic duals.
Finally, long quiver type theories have been studied in various other dimensions, e.g.\ recently in \cite{Nunez:2019gbg,Lozano:2019jza,Lozano:2019zvg,vanGorsel:2019sbz,Apruzzi:2017nck,Filippas:2019puw}, and it would be interesting to explore possible connections.

\begin{acknowledgments}
It is a pleasure to thank Benjamin Assel, Oren Bergman, Andrea Chaney, Martin Fluder, James Liu, Leopoldo Pando-Zayas and Alessandro Tomasiello for insightful discussions.
I am grateful to the organizers and participants of the workshop ``Holography, Generalized Geometry and Duality'' at the Mainz Institute for Theoretical Physics for the inspiring workshop, and to MITP for hospitality.
This work was supported, in part, by the US Department of Energy under Grant No.~DE-SC0007859,
by the Leinweber Center for Theoretical Physics,
and by the Mani L. Bhaumik Institute for Theoretical Physics.
\end{acknowledgments}

\appendix
\renewcommand\theequation{\Alph{section}.\arabic{equation}}

\section{Explicit \texorpdfstring{$\cA_\pm$}{Ap/Am} with D7 punctures}\label{app:Apm-expl}

Explicit expressions are derived for the functions $\cA_\pm$ and the regularity conditions for solutions with D7-brane punctures
(they were given in integral representations in \cite{DHoker:2017zwj}).
For $\Sigma$ the upper half plane, the functions $\cA_\pm$ for solutions with D7-branes can be written as 
\begin{align}
 \cA_\pm&= \cA_\pm^0 +  \sum_{\ell=1}^L \left[Z_\pm^\ell \ln(w-r_\ell)
   +Y^\ell \sum _{i=1}^I \frac{n_i^2}{4\pi} \cF_{i}(w,r_\ell)\right],
\nonumber\\
\cF_{i}(w,r_\ell)&=\int_{w_0}^w   \frac{dz}{z-r_\ell}\ln \left ( \gamma_i\,\frac{ z-w_i}{z -\bar w_i} \right ).
\end{align}
where $Y^\ell\equiv Z_+^\ell-Z_-^\ell$ and $\overline{Z_\pm^\ell}=-Z_\mp^\ell$.
The integration contour in $\cF_{i}(w,r_\ell)$ is to be chosen such that it does not cross the branch cut of the $\ln(\cdot)$ factor.
To evaluate $\cF_{i}(w,r_\ell)$ for fixed $i$, it is convenient to map the upper half plane to the unit disc, via
\begin{align}
 u&=\gamma_i\frac{w-w_i}{w-\bar w_i}~.
\end{align}
In the $u$ coordinate the branch cut is along the negative real axis, and  
\begin{align}
 \cF_i(w,r_\ell)&=\tilde \cF_i\left(\gamma_i\frac{w-w_i}{w-\bar w_i},\gamma_i\frac{r_\ell-w_i}{r_\ell-\bar w_i}\right),
 &
 \tilde \cF_i(u,t)&=\int_{u_0}^u d\tilde u\,  \ln(\tilde u)\left[\frac{1}{\tilde u-t}-\frac{1}{\tilde u-\gamma_i}\right].
\end{align}
In $\tilde \cF_i(u,t)$ dependence on $r_\ell$ is encoded in $t$, while $w_i$ and $\gamma_i$ are encoded in $u$ and $t$.
The second term in the square brackets does thus not depend on $\ell$. Since $\sum_\ell Y^\ell=0$ its contribution to $\cA_\pm$ vanishes. The remaining first term does not have explicit dependence on $\gamma_i$ or $w_i$.
This leaves
\begin{align}
 \cA_\pm&= \cA_\pm^0 +  \sum_{\ell=1}^L \left[Z_\pm^\ell \big[\ln(z-r_\ell)\big]_{w_0}^w
   +Y^\ell \sum _{i=1}^I \frac{n_i^2}{4\pi} \tilde L\left(\gamma_i\frac{w-w_i}{w-\bar w_i},\gamma_i\frac{r_\ell-w_i}{r_\ell-\bar w_i}\right)\right],
\nonumber\\
 \tilde L(u,t)&=\int_{u_0}^u d\tilde u\,  \frac{\ln(\tilde u)}{\tilde u-t_\ell}
 =\left[\Li_2\left(\frac{\tilde u}{t}\right)+\ln(\tilde u)\ln\left(1-\frac{\tilde u}{t}\right)\right]_{u_0}^u~.
\end{align}
With the branch cut of $\Li_2(z)$ extending from $z=1$ along the positive real axis and that of $\ln(z)$ extending from $z=0$ along the negative real axis, the explicit expression for $\tilde L$ has the correct analytic structure.
The contribution to $\cA_\pm$ from the lower integration bound can be absorbed into a redefinition of $\cA^0_\pm$,
combined with a shift as in (\ref{eq:cApm-SU11}) to maintain $\overline{\cA_\pm^0}=-\cA_\mp^0$. 
This leads to
\begin{align}
 \cA_\pm&= \cA_\pm^0 +  \sum_{\ell=1}^L \left[Z_\pm^\ell \ln(w-r_\ell)
   +Y^\ell \sum _{i=1}^I \frac{n_i^2}{4\pi} L\left(\gamma_i\frac{w-w_i}{w-\bar w_i},\gamma_i\frac{r_\ell-w_i}{r_\ell-\bar w_i}\right)\right],
 \nonumber\\
 L(u,t)&=\Li_2\left(\frac{u}{t}\right)+\ln(u)\ln\left(1-\frac{u}{t}\right)~.
 \label{eq:Apm-D7-app}
\end{align}

\tocless\subsection{Regularity conditions}

The remaining regularity conditions constraining the parameters stem from the requirement that $\cG$ has to vanish on the boundary of $\Sigma$ for regular solutions, and that the monodromy of $\partial_w\cA_\pm$ should lift properly to $\cA_\pm$.
They can be derived from
\begin{subequations}\label{eq:reg-D7-gen}
\begin{align}
\lim_{w\rightarrow w_i}\left[\cA_+(w)-\cA_-(w)+\rm{c.c.}\right]&=0~, & i&=i,\ldots,I~,
\label{eq:reg-monodromy}
\\
\lim_{w\rightarrow r_k}\left[
\cY_-^k\cA_+(w) -\cY_+^k\cA_-(w)
 -\rm{c.c.}\right]&=0~, & k&=1,\ldots,L~,
 \label{eq:reg-DeltaG}
\end{align}
\end{subequations}
where $\cY_\pm^k$ denotes the residue of $\partial_w\cA_\pm$ at $w=r_k$, as given in (\ref{eq:cY}).
In both conditions divergences cancel between the individual terms in the square brackets.
Evaluating these conditions with the expression for $\cA_\pm$ in (\ref{eq:Apm-D7-app}) leads to
\begin{subequations}
\begin{align}
\label{eq:app-reg-3-1}
  0&=2\cA_+^0-2\cA_-^0+\sum_{\ell=1}^LY^\ell \ln|w_i-r_\ell|^2~, & i&=1,\ldots,I~,
  \\
  0&=2\cA_+^0 Z_-^k-2\cA_-^0Z_+^k+\sum_{\ell\neq k}Z^{[\ell,k]}\ln|r_\ell-r_k|^2
  -Y^k\sum_{i=1}^I\frac{n_i^2}{4\pi} J_{ik}~, & k&=1,\ldots L~,
  \label{eq:app-reg-3-2}
\end{align}
\end{subequations}
where
\begin{align}\label{eq:Jik-def}
 J_{ik}&=\sum_{\ell=1}^L Y^\ell \lim_{w\rightarrow r_k}\left[L\left(\gamma_i\frac{w-w_i}{w-\bar w_i},\gamma_i\frac{r_\ell-w_i}{r_\ell-\bar w_i}\right)
 +\ln\left(\gamma_i\frac{r_k-w_i}{r_k-\bar w_i}\right)\left(\cA_+(w)-\cA_-(w)\right)
 -\rm{c.c.}
 \right].
\end{align}
Using the conditions in (\ref{eq:app-reg-3-1}) shows
\begin{align}
 \cA_+(w)-\cA_-(w)&=\sum_{\ell=1}^L Y^\ell \ln\frac{w-r_\ell}{|w_i-r_\ell|}~.
\end{align}
Using further the definition of $L$ in (\ref{eq:Apm-D7-app}) shows that the divergent terms in (\ref{eq:Jik-def}) cancel, and
\begin{align}
 J_{ik}&=\sum_{\ell=1}^L Y^\ell\left[
 \left(\Li_2\left(\frac{r_k-w_i}{r_k-\bar w_i}\frac{r_\ell-\bar w_i}{r_\ell-w_i}\right)-\rm{c.c.}\right)
 +\ln\left(\gamma_i\frac{r_k-w_i}{r_k-\bar w_i}\right)\ln\left|\frac{w_i-\bar w_i}{r_k-\bar w_i}\right|^2
 \right].
\end{align}
The second term does not depend on $\ell$, and drops out in the sum due to $\sum_\ell Y^\ell=0$.
This leads to the regularity conditions as given in (\ref{eq:reg-D7}) with (\ref{eq:cI-def}).
Since the remaining first term in $J_{ik}$ is antisymmetric under $k\leftrightarrow \ell$, the sum over the conditions in (\ref{eq:app-reg-3-2}) vanishes (note that the conditions in (\ref{eq:app-reg-3-1}) were used to derive this form of $J_{ik}$).

\tocless\subsection{Expressions with \texorpdfstring{$\ZZ_2$}{Z2} symmetry}
\label{eq:app-O7}

The expression for $\cA_\pm$ with $\ZZ_2$ symmetry can be obtained by splitting the sums over $\ell$ and $i$ in (\ref{eq:cApm-D7}) and using (\ref{eq:sol-O7-1}) and (\ref{eq:sol-O7-2}). This leads to
\begin{align}
 \cA_\pm(w)=\,&\cA_\pm^0+\sum_{\ell=1}^{L_0}\left[Z_\pm^\ell \ln\left(-i\frac{w-r_\ell}{wr_\ell+1}\right)+Y^\ell \sum _{i=1}^{I_0} \frac{n_i^2}{4\pi} \left(\tilde L_{i\ell}(w)-\tilde L_{i\ell}\left(-w^{-1}\right)\right)\right]~,
 \nonumber\\
\tilde L_{i\ell}(w)=\,&
L\left(\gamma_i\frac{w-w_i}{w-\bar w_i},\gamma_i\frac{r_\ell-w_i}{r_\ell-\bar w_i}\right)
+L\left(\gamma_i\frac{w w_i+1}{w \bar w_i+1},\gamma_i\frac{r_\ell w_i+1}{r_\ell \bar w_i+1}\right)~,
\end{align}
Constants arising from rewriting the $Z_\pm^\ell$ terms have been absorbed into a redefinition of $\cA_\pm^0$.
The branch cuts in the logarithmic terms connect $r_\ell$ and $-1/r_\ell$ through a half circle in the lower half plane, which is invariant under $w\rightarrow -1/w$.
The terms in the sum are odd under $w\rightarrow -1/w$, such that
\begin{align}
 \cA_\pm(w)+\cA_\pm(-1/w)&=2\cA_\pm^0~.
\end{align}
Taking the sum of the conditions in (\ref{eq:reg-DeltaG}) for $\ell$ and $\ell+L_0$ thus leads to
\begin{align}
 2\cY_-^k\left(\cA_+^0-\overline{\cA_-^0}\right)-2\cY_+^k\left(\overline{\cA_+^0}-\cA_-^0\right)&=0~, & \ell=1,\ldots,L_0~.
\end{align}
This fixes $\cA_+^0=\overline{\cA_-^0}$. One can then use the shifts in (\ref{eq:cApm-SU11}) to set, without loss of generality (up to a gauge transformation on $\cC$),
\begin{align}
 \cA_\pm^0&=0~.
\end{align}
The remaining regularity conditions evaluated from (\ref{eq:reg-D7-gen}) are
\begin{subequations}
\begin{align}
 0&=\sum_{\ell=1}^L Y^\ell \ln\left|\frac{w_i-r_\ell}{w_i r_\ell+1}\right|^2~, & i=1,\ldots,I_0~,
 \label{eq:O7-reg-0-1}
 \\
 0&=\sum_{\stackrel{\ell=1}{\ell\neq k}}^{L_0}\left[ Z^{[\ell,k]}\ln\left|\frac{r_\ell-r_k}{r_kr_\ell+1}\right|^2+Y^k Y^\ell \sum_{i=1}^{I_0}\frac{n_i^2}{4\pi}J_{ik\ell}\right]~, & \ell=1,\ldots,L_0~,
 \label{eq:O7-reg-0-2}
\end{align}
\end{subequations}
where, analogously to the general D7-puncture case, using the conditions in (\ref{eq:O7-reg-0-1}) and $\sum_\ell Y^\ell=0$, 
\begin{align}
 J_{ik\ell}=\,&
 \Li_2\left(\frac{r_k-w_i}{r_k-\bar w_i}\frac{r_\ell-\bar w_i}{r_\ell-w_i}\right)
 - \Li_2\left(\frac{r_k-w_i}{r_k-\bar w_i}\frac{r_\ell\bar w_i+1}{r_\ell w_i+1}\right)
 \nonumber\\ &
 +\Li_2\left(\frac{r_k w_i+1}{r_k \bar w_i+1}\frac{r_\ell\bar w_i+1}{r_\ell w_i+1}\right)
 -\Li_2\left(\frac{r_k w_i+1}{r_k \bar w_i+1}\frac{r_\ell-\bar w_i}{r_\ell-w_i}\right)-\rm{c.c.}
\end{align}

\bibliography{ads6O7}
\end{document}